\PassOptionsToPackage{table,xcdraw}{xcolor}
\documentclass[sigconf,authorversion,nonacm]{acmart}

\AtBeginDocument{%
  \providecommand\BibTeX{{%
    \normalfont B\kern-0.5em{\scshape i\kern-0.25em b}\kern-0.8em\TeX}}}

\setcopyright{acmcopyright}
\copyrightyear{2024}
\acmYear{2024}
\setcopyright{rightsretained}
\acmConference[CHI '24]{Proceedings of the CHI Conference on Human Factors in Computing Systems}{May 11--16, 2024}{Honolulu, HI, USA}
\acmBooktitle{Proceedings of the CHI Conference on Human Factors in Computing Systems (CHI '24), May 11--16, 2024, Honolulu, HI, USA}
\acmDOI{10.1145/3613904.3642089}
\acmISBN{979-8-4007-0330-0/24/05}

\usepackage{booktabs}
\usepackage{graphicx}
\usepackage[table,xcdraw]{xcolor}

\usepackage{multicol}
\usepackage{amsmath}
\usepackage{dsfont}
\usepackage{algorithm}
\usepackage{algpseudocode}
\usepackage{newtxmath}
\usepackage{ragged2e}
\usepackage[hang,flushmargin,bottom]{footmisc}

\usepackage{dirtytalk} % for quotations \say{something}
\usepackage{wrapfig}
\usepackage{subcaption}
\usepackage{multirow}
\usepackage[bottom]{footmisc}
\usepackage[nameinlink]{cleveref}
\usepackage{float}

\definecolor{myBlue}{RGB}{0, 122, 255}
\definecolor{myBlack}{RGB}{0, 0, 0}
\newcommand{\rev}[1]{{\color{myBlack}{#1}}}
\newcommand{\figcolor}[1]{{\color{myBlack}{#1}}}
\begin{document}

\title[LabelAId]{LabelAId: Just-in-time AI Interventions for Improving \\ Human Labeling Quality and Domain Knowledge \\ in Crowdsourcing Systems}


\author{Chu Li}
\authornote{Both authors contributed equally to this research.}
\email{chuchuli@cs.washington.edu}
\author{Zhihan Zhang}
\authornotemark[1]
\email{zzhihan@cs.washington.edu}
\affiliation{%
  \institution{University of Washington, USA}
  \country{}
}

\author{Michael Saugstad}
\affiliation{%
  \institution{University of Washington, USA}
  \country{}
}
\email{saugstad@cs.washington.edu}

\author{Esteban Safranchik}
\affiliation{%
  \institution{University of Washington, USA}
  \country{}
}
\email{estebans@cs.washington.edu}

\author{Minchu Kulkarni}
\affiliation{%
  \institution{University of Washington, USA}
  \country{}
}
\email{minchu@uw.edu}

\author{Xiaoyu Huang}
\affiliation{%
  \institution{University of California, Berkeley, USA}
  \country{}
}
\email{haytham.huang@berkeley.edu}

\author{Shwetak Patel}
\affiliation{%
  \institution{University of Washington, USA}
  \country{}
}
\email{shwetak@cs.washington.edu}

\author{Vikram Iyer}
\affiliation{%
  \institution{University of Washington, USA}
  \country{}
}
\email{vsiyer@cs.washington.edu}

\author{Tim Althoff}
\affiliation{%
  \institution{University of Washington, USA}
  \country{}
}
\email{althoff@cs.washington.edu}

\author{Jon E. Froehlich}
\affiliation{%
  \institution{University of Washington, USA}
  \country{}
}
\email{jonf@cs.washington.edu}

\renewcommand{\shortauthors}{Li et al.}

\begin{abstract}
  Crowdsourcing platforms have transformed distributed problem-solving, yet quality control remains a persistent challenge. Traditional quality control measures, such as prescreening workers and refining instructions, often focus solely on optimizing economic output. This paper explores just-in-time AI interventions to enhance both labeling quality and domain-specific knowledge among crowdworkers. We introduce LabelAId, an advanced inference model combining \textit{Programmatic Weak Supervision} (PWS) with \textit{FT-Transformers} to infer label correctness based on user behavior and domain knowledge. Our technical evaluation shows that our LabelAId pipeline consistently outperforms state-of-the-art ML baselines, improving mistake inference accuracy by 36.7\% with 50 downstream samples. We then implemented LabelAId into Project Sidewalk, an open-source crowdsourcing platform for urban accessibility. A between-subjects study with 34 participants demonstrates that LabelAId significantly enhances label precision without compromising efficiency while also increasing labeler confidence.
We discuss LabelAId's success factors, limitations, and its generalizability to other crowdsourced science domains.

\end{abstract}



\begin{teaserfigure}
  \includegraphics[width=\textwidth]{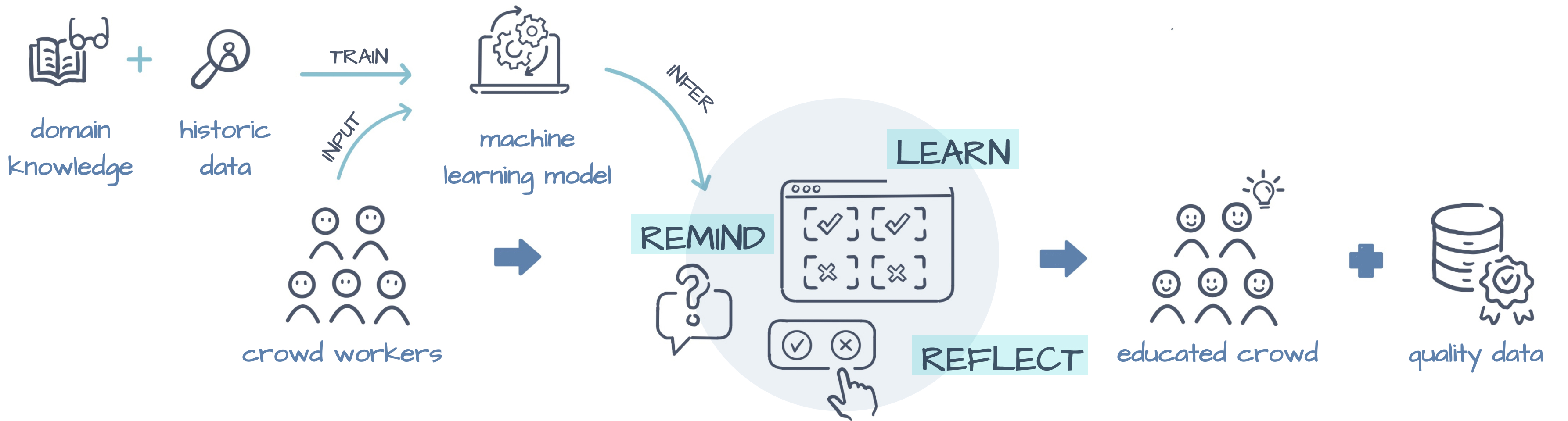}
  \caption{We introduce \textit{LabelAId}, an ML-based inference system to provide just-in-time feedback during crowdsourced labeling to improve data quality and user expertise. LabelAId consists of: (1) a novel ML-based pipeline for detecting labeling mistakes, which is efficiently trained to infer label correctness based on user behavior and domain knowledge; (2) a real-time ML model and UI that tracks worker behavior and intervenes when an inferred mistake is occurring.}
  \Description{An overview diagram of the LabelAId system, which contains one, a novel ML-based pipeline for detecting labeling mistakes, which is efficiently trained to infer label correctness based on user behavior and domain knowledge; and two, a real-time ML model and UI that tracks worker behavior and intervenes when an inferred mistake is occurring. LabelAId improves data quality and worker expertise by providing just-in-time feedback during crowdsource labeling.}
  \label{fig:teaser}
\end{teaserfigure}


\maketitle

%%%% the fun begins here:
\section{Introduction}

Crowdsourcing systems have transformed distributed human \\
problem-solving, enabling large-scale collaborations that were previously infeasible~\cite{ipeirotis_analyzing_2010}. 
Quality control, however, remains a persistent challenge leading to noisy or unusable data~\cite{kittur_future_2013, chen_cicero_2019}.
Existing quality control measures such as prescreening crowdworkers~\cite{difallah_pick--crowd_2013,kamar_combining_2012}, refining instructions~\cite{downs_are_2010, kittur_crowdsourcing_2008,shaw_designing_2011}, manipulating incentives~\cite{downs_are_2010, kittur_crowdsourcing_2008,shaw_designing_2011}, and majority vote filtering are designed to optimize economic output: data quality and worker efficiency.
Our research explores a subset of crowdsourcing that focuses on community science, or \textit{crowdsourced science}~\cite{reeves_crowd_2017}.
Platforms like \textit{Zooniverse}~\cite{simpson_zooniverse_2014} and \textit{FoldIt}~\cite{kelly_harnessing_2015} engage non-professionals in scientific tasks and serve as important means of public engagement and education~\cite{wang_exploring_2018, reeves_crowd_2017}.
Since participants are primarily volunteers, crowdsourced science presents unique quality control challenges: users are primarily motivated by intrinsic interest, learning opportunities, and making a difference but may be unfamiliar with the domain~\cite{phillips_framework_2018}.
Previous work in crowdsourcing has explored the dual objectives of enhancing work quality as well as learning experience in crowdsourcing systems by providing feedback to crowdworkers~\cite{dontcheva_combining_2014,doroudi_toward_2016,dow_shepherding_2012, wang_exploring_2018,zhu_reviewing_2014}. 
Yet, these approaches are less scalable because they require additional commitments from either crowdworker peers or external experts \cite{doroudi_toward_2016,dow_shepherding_2012, wang_exploring_2018,zhu_reviewing_2014}. 

Building on this prior work, we present \textit{LabelAId}, a real-time inference model for providing just-in-time feedback during crowdsource labeling to improve data quality and worker expertise. LabelAId is composed of two parts: (1) a novel machine learning (ML) based pipeline for detecting labeling mistakes, which is efficiently trained on unannotated data that contain those very mistakes; (2) a real-time system that tracks worker behavior and intervenes when an inferred mistake occurs. Unlike previous approaches that improve crowdworkers’ learning experience through peer or expert feedback~\cite{dow_shepherding_2012,zhu_reviewing_2014}, LabelAId reduces the reliance on human input, leveraging human-AI collaboration to provide targeted feedback for enhancing crowdworker performance and domain knowledge.

To study LabelAId in a real crowdsourcing context, we instrumented the open-source crowdsourcing tool, \textit{Project Sidewalk}, where online users virtually explore streetscape imagery to find, label, and assess sidewalk accessibility problems for people with mobility disabilities~\cite{saha_project_2019}.
Since its launch in 2015, over 13,000 people across the world have used Project Sidewalk to audit 17,000 $km$ of streets across 20 cities in eight countries including the US, Mexico, Ecuador, Switzerland, New Zealand, and Taiwan, contributing over 1.5 million data points\footnote{\noindent\href{https://projectsidewalk.org/}{https://projectsidewalk.org/}}. 

Project Sidewalk provides a compelling use case for LabelAId because, unlike traditional image labeling tasks for object detection (\textit{e.g.}, \textit{ImageNet}~\cite{deng_imagenet_2009}, \textit{COCO}~\cite{lin_microsoft_2015}, \textit{Open Images Dataset}~\cite{google_open_2022}), crowdworkers are asked to make careful judgments about a labeling target, which requires domain knowledge and training—similar to agricultural image recognition~\cite{ghosal_weakly_2019}, medical imagery labeling~\cite{radsch_labelling_2023,zhang_samdsk_2023}, and wildlife image categorization~\cite{berger-wolf_wildbook_2017}. Such labeling tasks reflect a broader trend of crowdwork becoming increasingly complex, domain-specific, and potentially error prone~\cite{kittur_future_2013}. Second, as a community science project, Project Sidewalk aligns with the growing emphasis on both educational impact and data quality in crowdsourcing~\cite{dontcheva_combining_2014,doroudi_toward_2016,dow_shepherding_2012, wang_exploring_2018,zhu_reviewing_2014}, which LabelAId provides. Finally, Project Sidewalk currently employs a common but limited quality control mechanism: users validate labeled images by other users. Since both labelers and validators are drawn from the same user population, repeated errors can pervade the system.

To evaluate LabelAId, we conducted: (1) a technical performance evaluation of LabelAId’s inference model; and (2) a between-subjects user study of 34 participants. For the former, we demonstrate that the LabelAId pipeline consistently outperforms state-of-the-art baselines and can improve mistake inference accuracy by up to 36.7\%. With fine-tuning on as few as 50 expert-validated labels, LabelAId outperforms traditional ML models such as \textit{XGBoost}~\cite{chen_xgboost_2016} and \textit{Multi-layer Perceptron }(MLP)~\cite{tolstikhin_mlp-mixer_2021} trained on 20 times the amount of expert-validated labels. Furthermore, we showcase the robust generalizability of our pipeline across different deployment cities in Project Sidewalk. Since its initial deployment in Washington D.C., Project Sidewalk has expanded to 20 cities, with ongoing plans for further growth. To support future city deployments, it is important to minimize labor and configuration overhead of the mistake inference model in new cities. Our study shows that LabelAId, even without fine-tuning, performs comparably in a new city to those in the pre-training set.

For the between-subjects user study, participants were randomly assigned to one of two conditions: using Project Sidewalk in its original form (control) or using Project Sidewalk with LabelAId (intervention). Our findings reveal that the intervention group achieved significantly higher label precision without sacrificing labeling speed. 
While using Project Sidewalk enhanced participants' understanding of urban accessibility and their confidence in identifying sidewalk problems in both groups, participants in the intervention group reported that LabelAId was helpful with decision-making, particularly in situations where they were initially uncertain. 

To summarize, our contributions are as follows:
\begin{itemize}
\item A novel ML pipeline that allows for the integration of domain-specific knowledge and heuristics into the data annotation process, which facilitates the training of AI-based inference models for detecting crowdworker labeling mistakes across various contexts, while minimizing the need for manual intervention in downstream tasks.
\item A human-AI (HAI) collaborative system designed to create teachable moments in crowdsourcing workflows. This system not only improves the quality of crowdsourced data, but also enriches the learning experience for participants.
\item A between-subjects user study involving 34 participants with no prior experience using Project Sidewalk, demonstrating that LabelAId significantly improves label precision by 19.2\% without compromising efficiency.
\end{itemize}

While our empirical results focused on the performance of LabelAId within the context of Project Sidewalk, we believe our framework can be generalizable to other crowdsourcing platforms as well as the PWS-based ML pipeline and the two-step module design intervention are easily replicable and tailorable in different contexts. 
\section{Related Work}
Our work draws on, and contributes to research in improving the quality of crowdsourcing, enhancing crowdworkers' domain knowledge, and inferring the correctness of labels using ML methods.

\begin{figure*}[t]
  \centering
  \includegraphics[width=\linewidth]{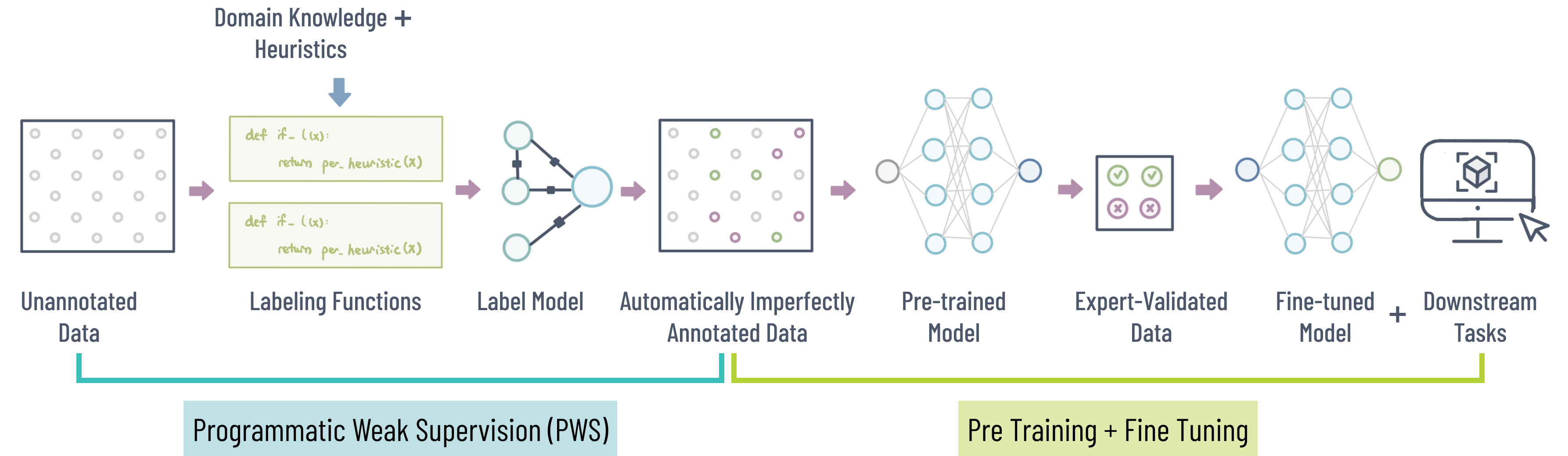}
  \caption{An overview of our LabelAId pipeline. Programmatic weak supervision, utilizing domain-specific knowledge and heuristics, is employed to annotate the raw data. Subsequently, the automatically imperfectly annotated data generated from PWS are used to pre-train the inference model. Lastly, the inference model is fine-tuned using expert-validated labels for the target downstream task. Diagram adapted from~\cite{ratner_snorkel_2017}.}
  \Description{Conceptual diagram of our LabelAId pipeline. Programmatic weak supervision, using domain-specific knowledge and heuristics, is employed to annotate the raw data. Subsequently, the automatically imperfectly annotated data generated from PWS are used to pre-train the inference model. Lastly, the inference model is fine-tuned using expert-validated labels for the target downstream task.}
  \label{fig:pipeline}
\end{figure*}

\subsection{Improving Quality of Crowdsourced Labels}
Distributed crowdwork has transformed how loosely connected individuals collaborate together to solve large-scale problems such as protein folding~\cite{cooper_predicting_2010} map building~\cite{haklay_citizen_2013}, and writing compendiums of knowledge~\cite{wilkinson_cooperation_2007}. Despite decades of research, however, large-scale crowdwork remains susceptible to quality control problems~\cite{chang_revolt_2017, kittur_future_2013}. For example, studies have shown that over 30\% of MTurk submissions are likely to be poor quality~\cite{bernstein_soylent_2010, kittur_crowdsourcing_2008}. Current quality control methods can be broadly categorized into two groups: preventive techniques and post-hoc detections. Preventive measures include screening crowdworkers based on capabilities~\cite{difallah_pick--crowd_2013,kamar_combining_2012}, dividing work into fault-tolerant sub-tasks~\cite{bernstein_soylent_2010,kittur_crowdforge_2011,noronha_platemate_2011}, improving instructions~\cite{downs_are_2010, kittur_crowdsourcing_2008,shaw_designing_2011} and changing payment structures~\cite{chandler_breaking_2013, rogstadius_assessment_2011, shaw_designing_2011}. Post-hoc measures involve filtering based on majority vote~\cite{snow_cheap_2008} and employing additional crowdworkers to review others' work~\cite{callison-burch_creating_2010, hansen_quality_2013}. Project Sidewalk currently uses both strategies: an interactive tutorial to train crowdworkers as their ``first mission'' and \textit{post-hoc} validation where crowdworkers ``vote'' on the correctness of other users' labels.

Other quality control research examines how workers do their work rather than the end product itself, using ML algorithms to predict the quality of crowdworkers’ output based on their behaviors~\cite{rzeszotarski_crowdscape_2012,rzeszotarski_instrumenting_2011, gadiraju_crowd_2019}. 
This method captures behavioral traces from workers during task execution and uses them to predict quality, errors, and the possibility of cheating~\cite{rzeszotarski_crowdscape_2012,rzeszotarski_instrumenting_2011, gadiraju_crowd_2019}.
These behavioral traces are gathered by logging user interactions, which are then formulated into interaction patterns for monitoring real-time worker compliance ~\cite{rzeszotarski_instrumenting_2011}. 
This methodology, termed \textit{``fingerprinting''} by Rzeszotarski and Kittur~\cite{rzeszotarski_instrumenting_2011}, has demonstrated its efficacy in predicting crowdworker output quality.
Expanding Rzeszotarski and Kittur’s work, Kaza and Zitoun~\cite{kazai_quality_2016} investigated using the behavior of trusted, trained judges to identify low-performing workers. 
Their study, which involved assessing the relevance of web pages to specific queries, showed that the classification accuracy nearly doubled in some tasks. 
However, the approach is challenging to scale due to the need for trained judges.

Building on this body of research, we introduce an ML pipeline that combines crowdworker behavioral data with expert domain knowledge (in our case, drawn from urban accessibility but the approach should generalize to other domains). This model aims to more effectively and automatically guide crowdworkers through their efforts in identifying street-level accessibility issues.

\subsection{Teachable Moments in Crowdsourcing for Community Science}
Crowdsourcing for community science are initiatives where professional scientists seek the assistance of crowds in contributing to scientific research~\cite{haklay_citizen_2013,reeves_crowd_2017}. Platforms like \textit{Zooniverse}~\cite{simpson_zooniverse_2014}, \textit{FoldIt}~\cite{kelly_harnessing_2015}, and \textit{SciStarter}~\cite{hoffman_scistarter_2001} are notable for having involved non-professionals in significant scientific discoveries.
Beyond contributing to science, these platforms serve as tools for public engagement, outreach, and education~\cite{wang_exploring_2018}. 
Unlike crowdworkers driven by monetary incentives (\textit{e.g.,} on MTurk and Prolific), participants of community science projects are primarily volunteers motivated by desires to learn and contribute to scientific research~\cite{raddick_galaxy_2013, reed_exploratory_2013}. 

Recent crowdsourcing research has been investigating ways to not only enhance the quality of work but also the learning experience of participants~\cite{dontcheva_combining_2014,doroudi_toward_2016,dow_shepherding_2012, wang_exploring_2018,zhu_reviewing_2014}. For instance, Dow \textit{et al.}~\cite{dow_shepherding_2012} demonstrated that timely, task-specific feedback can help crowdworkers learn, preserve, and produce better results. Projects like \textit{Crowdclass}~\cite{lee_crowdclass_2016} and \textit{CrowdSCIM}~\cite{wang_exploring_2018} introduced in-task learning modules for community science initiatives. Despite these advances, current methods facilitating learning through crowdsourcing, such as peer-review~\cite{dow_shepherding_2012, zhu_reviewing_2014}, expert feedback~\cite{dow_shepherding_2012}, and self-assessment~\cite{dow_shepherding_2012}, all require additional commitments from either the crowdworkers or external experts, limiting their scalability.

Recent developments in HAI collaboration presents new ways to tackle these scalability issues. Matsubara \textit{et al.}~\cite{matsubara_learning_2018} suggest using machine predictions as reference answers for self-correction. Nakayama \textit{et al.}~\cite{nakayama_crowd-worker_2021} extended this concept, proposing workflows where AI learns alongside human workers without prior training. Inspired by this evolving landscape, we propose an add-on system for crowdsourcing systems like Project Sidewalk that enhances both task quality and educational outcomes. It leverages HAI collaboration to enable in-context learning without requiring additional commitments from participants.

\subsection{Machine Learning to Infer Label Correctness}
To create teachable moments in crowdsourcing workflows, it is essential to develop inference models for detecting crowdworker labeling mistakes. Recent research has harnessed the power of ML to infer crowdsourced label quality. For instance, computer vision-based neural networks are applied to validate crowdsourced labels of sidewalk accessibility problems in \textit{Google Street View} (GSV) imagery~\cite{weld_deep_2019}, and also offer reference answers to image recognition labelers enabling self-correction~\cite{matsubara_learning_2018, nakayama_crowd-worker_2021}. 
These deep learning methods present promising solutions for aiding crowdsourced labeling tasks, but they often require substantial training data. While general labeled image datasets such as ~\cite{deng_imagenet_2009, google_open_2022, lin_microsoft_2015} are available, domain-specific datasets are relatively rare and expensive to produce. 

In this paper, we aim to minimize the need for manually-annotated data for training AI-based inference models. 
The recently proposed \textit{Programmatic Weak Supervision} (PWS) framework~\cite{ratner_snorkel_2017, ratner_training_2019, ratner_data_2016} provides a promising approach: aggregating noisy votes from domain knowledge, heuristics, external patterns and rules to assign annotations to the raw data. Subsequently, this annotated data serves to train models for a range of domain-specific tasks, including video analysis~\cite{varma_multi-resolution_2019}, text classification~\cite{shu_learning_2020}, and sensor data analysis~\cite{furst_applying_2020}. However, its potential for inferring and improving the quality of crowdsourced labels remains unexplored.
\begin{table*}[t]
\resizebox{0.85\linewidth}{!}{%
\begin{tabular}{@{}
>{\columncolor[HTML]{FFFFFF}}l l
>{\columncolor[HTML]{FFFFFF}}r 
>{\columncolor[HTML]{FFFFFF}}r l
>{\columncolor[HTML]{FFFFFF}}r 
>{\columncolor[HTML]{FFFFFF}}r l
>{\columncolor[HTML]{FFFFFF}}r 
>{\columncolor[HTML]{FFFFFF}}r l
>{\columncolor[HTML]{FFFFFF}}r @{}}
\toprule
\cellcolor[HTML]{FFFFFF} &
   &
  \multicolumn{2}{c}{\cellcolor[HTML]{FFFFFF}Seattle} &
   &
  \multicolumn{2}{c}{\cellcolor[HTML]{FFFFFF}Chicago} &
   &
  \multicolumn{2}{c}{\cellcolor[HTML]{FFFFFF}Oradell} &
   &
  \cellcolor[HTML]{FFFFFF} \\ \cmidrule(lr){2-11}
\multirow{-2}{*}{\cellcolor[HTML]{FFFFFF}} &
   &
  Unannotated &
  \begin{tabular}[c]{@{}r@{}}Expert-\\ Validated\end{tabular} &
   &
  \multicolumn{1}{c}{\cellcolor[HTML]{FFFFFF}Unannotated} &
  \multicolumn{1}{c}{\cellcolor[HTML]{FFFFFF}\begin{tabular}[c]{@{}c@{}}Expert-\\ Validated\end{tabular}} &
   &
  \multicolumn{1}{c}{\cellcolor[HTML]{FFFFFF}Unannotated} &
  \multicolumn{1}{c}{\cellcolor[HTML]{FFFFFF}\begin{tabular}[c]{@{}c@{}}Expert-\\ Validated\end{tabular}} &
   &
  \multirow{-2}{*}{\cellcolor[HTML]{FFFFFF}\textbf{Total}} \\
\cellcolor[HTML]{F3F3F3}Curb Ramp &
  \cellcolor[HTML]{F3F3F3} &
  \cellcolor[HTML]{F3F3F3}70,690 &
  \cellcolor[HTML]{F3F3F3}5,333 &
  \cellcolor[HTML]{F3F3F3} &
  \cellcolor[HTML]{F3F3F3}5,710 &
  \cellcolor[HTML]{F3F3F3}2,386 &
  \cellcolor[HTML]{F3F3F3} &
  \cellcolor[HTML]{F3F3F3}660 &
  \cellcolor[HTML]{F3F3F3}859 &
  \cellcolor[HTML]{F3F3F3} &
  \cellcolor[HTML]{F3F3F3}\textbf{85,638} \\
Missing Curb Ramp &
   &
  32,968 &
  4,239 &
   &
  463 &
  1,294 &
   &
  325 &
  396 &
   &
  \textbf{39,685} \\
\cellcolor[HTML]{F3F3F3}No Sidewalk &
  \cellcolor[HTML]{F3F3F3} &
  \cellcolor[HTML]{F3F3F3}36,021 &
  \cellcolor[HTML]{F3F3F3}3,460 &
  \cellcolor[HTML]{F3F3F3} &
  \cellcolor[HTML]{F3F3F3}2,211 &
  \cellcolor[HTML]{F3F3F3}48 &
  \cellcolor[HTML]{F3F3F3} &
  \cellcolor[HTML]{F3F3F3}3,949 &
  \cellcolor[HTML]{F3F3F3}1,217 &
  \cellcolor[HTML]{F3F3F3} &
  \cellcolor[HTML]{F3F3F3}\textbf{46,906} \\
Surface Problem &
   &
  26,912 &
  2,909 &
   &
  2,136 &
  1,651 &
   &
  2,544 &
  1,222 &
   &
  \textbf{37,374} \\
\cellcolor[HTML]{F3F3F3}Obstacle &
  \cellcolor[HTML]{F3F3F3} &
  \cellcolor[HTML]{F3F3F3}10,103 &
  \cellcolor[HTML]{F3F3F3}407 &
  \cellcolor[HTML]{F3F3F3} &
  \cellcolor[HTML]{F3F3F3}1,254 &
  \cellcolor[HTML]{F3F3F3}320 &
  \cellcolor[HTML]{F3F3F3} &
  \cellcolor[HTML]{F3F3F3}106 &
  \cellcolor[HTML]{F3F3F3}158 &
  \cellcolor[HTML]{F3F3F3} &
  \cellcolor[HTML]{F3F3F3}\textbf{12,348} \\
\textbf{All Label Types} &
   &
  \textbf{176,694} &
  \textbf{16,348} &
   &
  \textbf{11,774} &
  \textbf{5,699} &
   &
  \textbf{7,584} &
  \textbf{3,852} &
   &
  \textbf{221,951} \\ \bottomrule
\end{tabular}%
}
\vspace{0.3cm}
\caption{Set sizes of unannotated and expert-validated labels, contain Project Sidewalk labels from Seattle, WA; Chicago, IL; and Oradell, NJ.
The unannotated set is used to pre-train the model after our PWS annotation process. 
The expert-validated set is used to fine-tune and evaluate the inference model, which was created from labels manually-validated by the Project Sidewalk research team.}
\label{tab:PS-dataset-size}
\vspace{-0.75cm}
\end{table*}

\section{LabelAId: A Label Correctness Inference Framework}
\label{sec:labelaid}

LabelAId is a novel ML pipeline designed to provide just-in-time intervention in crowdsourced labeling tasks, inferring and identifying labeling mistakes as they occur. At its core, LabelAId tackles a major hurdle in deploying such AI-based inference models: the need for large volumes of annotated training data, which is particularly scarce in crowdsourced environments. LabelAId introduces both (1) a programmatic pipeline to train an efficient ML inference model to detect crowdworker labeling mistakes, which is trained on unannotated data that contain those very mistakes and minimal expert-validated data, and (2) an example application of the LabelAId pipeline to a crowdsourcing system to recognize and intervene when a user is making a labeling mistake. Our overarching goal is to create a full end-to-end HAI pipeline with minimal expert involvement for model training, thereby facilitating rapid deployment of LabelAId mistake inference models in real-world crowdsourcing contexts. In this section, we describe LabelAId's pipeline design, how we implemented LabelAId in Project Sidewalk, and our technical evaluation. Our code is available on Github.\footnote{\href{https://github.com/makeabilitylab/LabelAId}{https://github.com/makeabilitylab/LabelAId}}

\subsection{LabelAId Pipeline}

Developing an inference model capable of discerning specific user mistakes while also being generalizable to general crowdworkers' behaviors is challenging and necessitates the consideration of a variety of quality signals. Our LabelAId pipeline is composed of three core phases (\autoref{fig:pipeline}): (1) \textit{Programmatic weak supervision} (PWS) uses domain-specific knowledge about crowdsourcing tasks and historical heuristics from crowdworker behavior; this phase generates a set of \textit{Automatically, Imperfectly Annotated} (AIA) probabilistic training data. (2) We then pre-train an ML model using the AIA data generated from PWS. The inference model learns general features and representations of the target crowdsourcing task in this phase. (3) Finally, we fine-tune the inference model using a small number of expert-validated labels to further enhance its performance for the target task.

\subsubsection{Programmatic Weak Supervision (PWS)}
Due to the inherently noisy nature of crowdsourced data, LabelAId adopts PWS as its foundational architecture. PWS takes unannotated data and produces probabilistic training labels, and has demonstrated effectiveness in tasks like document and numerical data classification~\cite{zhang_understanding_2022, bringer_osprey_2019}. We use the popular Snorkel~\cite{ratner_snorkel_2017} platform as the backbone of our PWS pipeline. PWS allows for the integration of domain knowledge and heuristic guidelines into the data annotation process and provides a method for estimating their conflict and correlation in a programmatic manner. As a result, the probabilistic training labels can be reweighted and combined to create high-quality labels. This approach aligns well with our objectives for two key reasons: first, it allows us to train models on unannotated data, eliminating the need for manually labeling large datasets; and second, it enables the incorporation of domain-specific knowledge related to crowdsourcing tasks into our pipeline.

One of the core components of Snorkel is \textit{Labeling Functions} (LFs), which are rules or heuristics humans code to annotate raw data programmatically. 
When incorporating a set of LFs, they introduce distinct characteristics and correlations that extend beyond disparities in accuracy and coverage. It is important to recognize that LFs are not equal in their contribution to the annotation. LFs may also overlap and conflict, which cannot be resolved by a simplistic hard-rule-based approach (see~\Cref{sec:pws-additional} for additional discussion). While previous approaches used majority-vote-based models to handle these intricacies, it can result in an overrepresentation of specific signals when two features are highly correlated~\cite{ratner_snorkel_2017}.
To address this, we instead opt for probabilistic graphical models to integrate the outputs of LFs. We use the Snorkel label model to take the complete set of LFs as its input and generate a matrix of LFs, $\Lambda$. We aim to maximize the probability of the outputs of the LFs~\cite{zhang_survey_2022} in the context of our label correctness inference task, \textit{i.e.}, a binary classification. Assuming $\theta$ is the set of model parameters and $Y$ is the prediction class, this objective transforms into an optimization problem as the following equation:

\[\underset{\theta}{\max}\ P(\Lambda; \theta)  = \sum_{Y}^{} P(\Lambda, Y; \theta) \]

The label model generates a single set of noise-aware, probabilistic training labels, which can be used to train an ML model or neural network.

\subsubsection{Transfer Learning from AIA to Expert-Validated Labels}
The ultimate goal is to train a discriminative ML model—such as a neural network classifier—that can generalize beyond the label model. The choice of the specific ML model architecture should be tailored to the requirements of the downstream task.

While PWS produces a set of AIA probabilistic training data, an ML model trained on a large yet noisy dataset could potentially overfit to the noise, which should be avoided. Fine-tuning large pre-trained models is a popular methodology for leveraging knowledge from a related source domain to improve the learning performance in a target task with limited clean samples~\cite{zhuang_comprehensive_2020}. 

Since the source domain and target task in our LabelAId pipeline share the same feature space, we introduce a two-phase transfer learning pipeline from AIA data to expert-validated labels: (1) Pre-training the ML model on the AIA dataset $\mathcal{D}_n$ produced by our PWS pipeline; (2) Fine-tuning the pre-trained model on a small number of expert-validated downstream labels $\mathcal{D}_g$ for a specific task. Our pre-training and fine-tuning pipeline is illustrated in \autoref{fig:pipeline}. This approach aims to minimize the requirement of manual intervention in annotating downstream data for a target task.

\subsection{Applying LabelAId to Project Sidewalk}
\begin{figure*}[t]
  \centering
  \includegraphics[width=\linewidth]{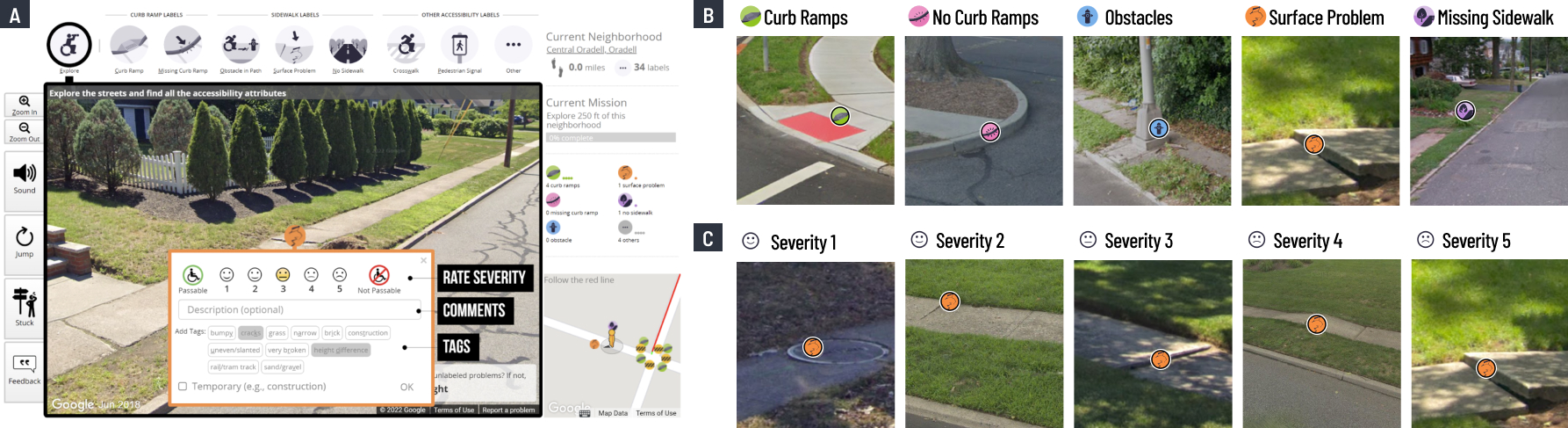}
  \caption{(A) Project Sidewalk Labeling Interface. (B) Project Sidewalk Label Types. (C) Examples of Project Sidewalk severity ratings for surface problems. Severity 5 is the most severe, indicating a scenario impassable by wheelchair users.}
  \Description{This figure showcases the Project Sidewalk Labeling Interface. Part A displays the interface used for labeling. Part B lists the different types of labels that can be applied, such as "obstacle" or "clear path." Part C provides examples of severity ratings for surface problems, ranging from 1 to 5, with 5 being the most severe and indicating a surface impassable for wheelchair users.}
  \label{fig:project-sidewalk}
\end{figure*}

We present a specific implementation of our end-to-end LabelAId pipeline using domain knowledge in urban accessibility for Project Sidewalk (\autoref{fig:project-sidewalk}), a large-scale and widely used sidewalk accessibility platform leveraging crowdsourced labels. We describe the dataset, details of labeling functions and features, and model training process below.

\subsubsection{Dataset Description}
At the time of our analysis, Project Sidewalk had 757,730 crowdsourced image labels applied on top of \textit{Google Street View} (GSV) panoramas across 15 cities in the US, Mexico, and Europe. Project Sidewalk has five main label types: \textit{Curb Ramp}, \textit{Missing Curb Ramp}, \textit{Sidewalk Obstacle}, \textit{Surface Problem}, \textit{Missing Sidewalk}.
Each label includes a severity assessment on a scale of 1 to 5, with 5 being the most severe, indicating a sidewalk issue that is impassable for a wheelchair user. Labels may also include an open-ended description and label-specific tags. Additionally, all labels are accompanied by implicitly captured metadata, such as the GSV image date, the label timestamp, and geographical location (latitude and longitude).

As urban composition and sidewalk accessibility guidelines differ across regions, our initial proof-of-concept of LabelAId focuses on three U.S. cities: Seattle, WA; Chicago, IL; and Oradell, NJ. These three cities offer distinct urban and geographical characteristics: Seattle represents a major city in the Pacific Northwest, the Chicago data provides a mix of dense downtown area and satellite towns in the Midwest, and Oradell is a suburban locale on the East Coast outside of New York City. We anticipate that this selection will help us develop ML models that could effectively account for diverse urban compositions. The ground truth dataset $\mathcal{D}_g$ is validated by the Project Sidewalk research team; two researchers worked collaboratively to verify each crowdsourced label, reaching unanimous agreement. The finalized dataset sizes alongside the distribution of each label type are shown in \autoref{tab:PS-dataset-size}.

\subsubsection{Input Features \& Labeling Functions}

The suitability of PWS for Project Sidewalk comes from two key factors: the integration of domain-specific knowledge (\textit{e.g.}, urban planning guidelines) and user behavior insights into our labeling process.

Drawing on observations of user behavior in Project Sidewalk and research in urban planning guidelines, we propose the following hypotheses:

\begin{itemize}

\item \textbf{Severity Rating.} Project Sidewalk's label severity ranges between 1 to 5. Severity ratings closer to the extremes (1 \& 5) are more likely to be correct.

\item \textbf{Optional input.} Labels that include optional data are more likely to be accurate, because such information requires additional thought and effort. These optional fields include a free-form description (comment) and relevant tags, such as \emph{fire hydrant} and \emph{pole} for the \emph{obstacle} label type. 

\item \textbf{GSV zoom/pitch/heading.} In most cases, changing the default parameters of GSV results in a more accurate label. For example, when a user zooms in to place a \textit{Surface Problem} label, it is more likely to be correct.

\item \textbf{Distance to other crowdworkers' labels.} A label is more likely to be correct if it is placed closer to existing labels of the same type. To determine this distance, we adopt the two-step spatial clustering approach employed in Project Sidewalk~\cite{saha_project_2019}.

\item \textbf{Distance to urban infrastructure.} The positioning of a label in relation to urban infrastructure can serve as an indicator of its accuracy. For example, US federal legislation~\cite{us_department_of_transportation_dot_curb_2023} requires the installation of curb ramps at all intersections and at midblock locations where pedestrian crossings are present. Given that midblock crossings are relatively rare compared to those at intersections, and considering that the most common error in Project Sidewalk is mislabeling driveways as curb ramps~\cite{weld_deep_2019}, we hypothesize that a \textit{Curb Ramp} label situated outside a specified radius from an intersection is likely to be incorrect.
\end{itemize}

\begin{algorithm}[b]
    \caption{Example Labeling Function encoding a heuristic about errors made in (Missing) Curb Ramp labels}
    \label{algorithm:algorithm-1}
    \begin{algorithmic}
    \Require \text{labels $\in$ CurbRamp or NoCurbRamp}
    \Ensure  \text{labels $\in$ Residential Area}
    \State $D \gets \text{Intersection Distance Threshold}$
        \For{each label $l$}
            \For{each intersection $i$ in nearby intersections $I$}
                \State \textbf{Compute} \text{spatial distance between $l$ and $i$}
            \EndFor
            
            \If{$\min \text{distance}(l, I) > D$}
                \State $l \gets \text{wrong}$
            \Else
                \State $l \gets \text{correct}$
            \EndIf
        \EndFor
    \end{algorithmic}
\end{algorithm}

We then derived eight LFs from our hypotheses for Project Sidewalk and integrated all these LFs into the PWS pipeline to ensure a diverse coverage and minimize overfitting (see~\Cref{app:lf-additional} for additional discussion).
One example algorithm (\autoref{algorithm:algorithm-1}) is based on the observation that users often mislabel driveways as curb ramps in residential areas. The algorithm proposes that a \textit{Curb Ramp} or \textit{Missing Curb Ramp} in a residential area is likely to be wrong when it is far away from an intersection.

\subsubsection{Multi-city Pre-training} 
\begin{figure*}[t]
  \centering
  \includegraphics[width=\linewidth]{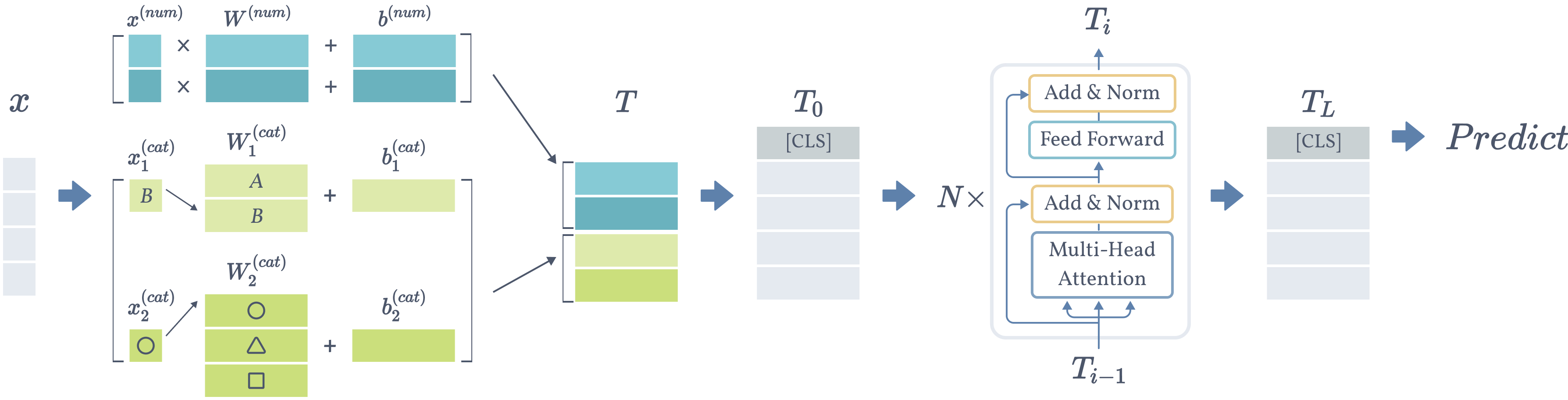}
  \caption{Conceptual diagram of our FT-Transformer-based model architecture. First, the model transforms the hybrid features (\textit{e.g.}, two numerical and two categorical features) into unified embeddings. Subsequently, these embeddings are processed iteratively by the Transformer layer. The final output is based on the [CLS] token. Diagram adapted from~\cite{gorishniy_revisiting_2021}.}
  
  \Description{Conceptual diagram of our FT-Transformer-based model architecture. First, the model transforms the hybrid features (e.g., two numerical and two categorical features) into unified embeddings. Subsequently, these embeddings are processed iteratively by the Transformer layer. The final output is based on the [CLS] token.}
  \label{fig:transformer-diagram}
\end{figure*}

To train a discriminative model, we start by pre-training on the AIA dataset $\mathcal{D}_n$ to initialize the weights for high-level patterns of sidewalk accessibility labels and user behaviors.
Due to the mix of categorical and numerical features within Project Sidewalk's datasets, we chose the \textit{Feature Tokenizer + Transformer} (FT-Transformer)~\cite{gorishniy_revisiting_2021}. FT-Transformer represents a novel adaptation of the \textit{Transformer} architecture for tabular data domains. Prior research~\cite{gorishniy_revisiting_2021,levin_transfer_2023} has shown that the FT-Transformer is a more universal architect for tabular data, and consistently outperforms other state-of-the-art deep tabular models across a variety of downstream sample regimes. A high-level view of our FT-Transformer-based model architecture is illustrated in \autoref{fig:transformer-diagram}. 

We employ three distinct datasets to train, validate, and test our backbone FT-Transformer-based model. We used the AIA datasets $\mathcal{D}_n$ from three cities generated from our PWS pipeline, which consists of 176,694, 11,774, and 7,584 labels for Seattle, Chicago, and Oradell, respectively. We balanced and randomly partitioned these labels into training, validation, and testing sets following a 70/20/10 split. 
In each subset, a minimum of 20 labels were guaranteed for every class of every label type.

We trained an FT-Transformer from scratch. To map all inputs into the same embedding space, we encode the numerical data through a single-layer perceptron with a dimension of four, and embed categorical data with one-hot embeddings of the same size. Then, we stack them to formulate the input embeddings for the Transformer module. The encoder was configured with a depth of two layers, each comprised of two attention heads. To prevent overfitting and to provide regularization, we incorporated attention and feed-forward dropouts, both set at 0.2. Additional optimal hyperparameters were tuned via grid search, according to the full upstream dataset: the \textit{AdamW} optimizer, a learning rate of $1 \times 10^{-4}$, and a weight decay of $1 \times 10^{-5}$ were selected. Given our binary classification objective, we employed the \textit{Binary Cross-Entropy} loss function coupled with a sigmoid activation function. We employed 200 epochs for training with early stops on validation loss. The average training duration was approximately 5 minutes on a single RTX 4070Ti GPU.

\subsubsection{Fine-tuning on a Specific City}
Upon delving deeper into our pipeline, the pre-trained model, which learned underlying patterns of sidewalk accessibility labels and a broad understanding of high-level user behaviors, can subsequently be fine-tuned on a smaller, city-specific $\mathcal{D}_g$ to elevate its performance for city-specific tasks. The rationale behind this is that each city has unique attributes that might not be entirely captured during this pre-training phase. Through fine-tuning, the model can adapt its previously acquired knowledge to the unique characteristics and nuances of the target city. This not only enables more precise inferences and understanding of the distinctive topologies of the target city, but also ensures that the model remains robust. To evaluate its performance enhancement, we conducted city-specific fine-tuning. The pre-trained FT-Transformer was fine-tuned end-to-end~\cite{levin_transfer_2023} with 200 downstream samples (40 per label type) sourced from Seattle, Chicago, and Oradell's $\mathcal{D}_g$, with a reduced learning rate of $3 \times 10^{-5}$ to prevent overfitting to downstream samples.

\subsection{Technical Evaluation}
We evaluate the performance of our pipeline, which involves the PWS pipeline integrated with city-specific fine-tuning of a pre-trained inference model. We examine the following aspects: (1) How our pipeline performs in comparison to traditional ML methods in inferring label correctness;
(2) The generalization of our model across cities after target-city-specific fine-tuning; 
(3) The model's generalizability to new cities that were not included in our pre-training dataset; 
(4) How different types of LFs complement each other in achieving LabelAId’s performance through analysis of feature importance.

For testing purposes, we utilized the remaining $\mathcal{D}_g$ after removing the samples used for fine-tuning. This allowed us to evaluate the performance and generalizability of the final model on a previously unseen dataset. Specifically, the remaining $\mathcal{D}_g$ comprises 16,148, 5,499, and 3,652 expert-validated labels for Seattle, Chicago, and Oradell, respectively.

\subsubsection{Pipeline Performance}
We compared our performance against two widely adopted ML classifiers–random forest and logistic regression, and a GBDT tabular method XGBoost~\cite{chen_xgboost_2016}. We also considered an MLP classifier which is known to be a consistent and competitive baseline~\cite{gorishniy_revisiting_2021}. All baselines are trained on the equivalent volumes of expert-validated downstream samples |$\mathcal{D}_g$| and do not undergo pre-training using $\mathcal{D}_n$ produced by our PWS pipeline.

To ensure a robust evaluation, the optimal hyperparameters for baseline models were tuned via a grid search, executed on a single, randomly split downstream dataset. This procedure ensures that all baselines are tuned with an equivalent number of samples, given that the hyperparameters are profoundly influenced by sample size. Each baseline model underwent training using 50, 100, 200, 500, and 1000 expert-validated labels from Seattle $\mathcal{D}_g$. The samples were equally distributed across two classes and five label types. For our experiment, we fine-tuned the pre-trained FT-Transformer end-to-end~\cite{levin_transfer_2023} on the equivalent number of expert-validated downstream samples from Seattle $\mathcal{D}_g$ with the baselines.

\begin{figure}[b]
  \centering
  \includegraphics[width=\linewidth]{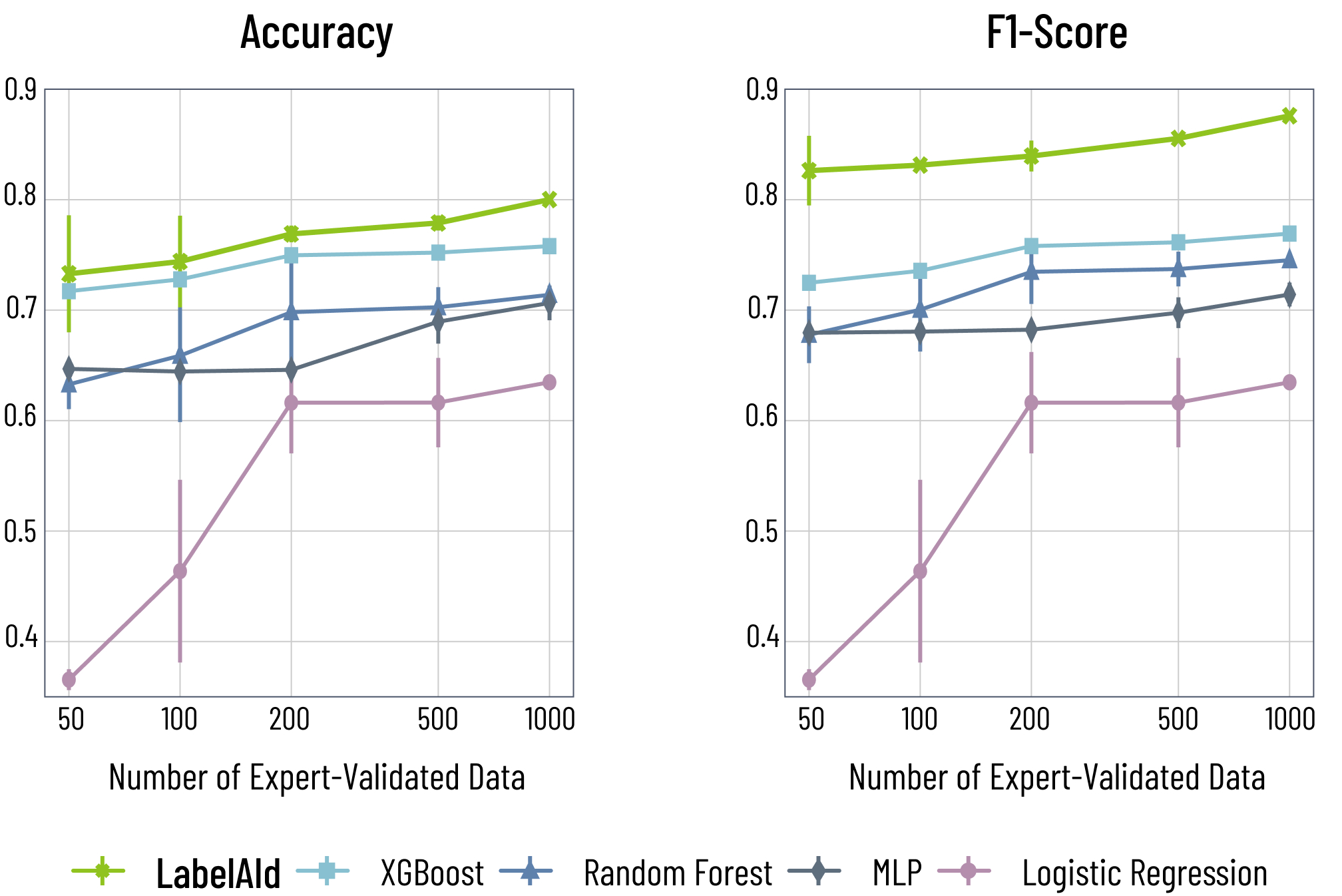}
  \caption{Overall performance of our LabelAId pipeline compared to the traditional ML methods as the number of expert-validated downstream labels increases. Note that the x-axis is on a log scale (N = 3, error bar = $\pm\sigma$).
  }
  \Description{Overall performance of our LabelAId pipeline compared to the traditional ML methods as the number of expert-validated downstream labels increases. Note that the x-axis is on a log scale (N = 3, error bar = $\pm\sigma$). Each baseline model is trained on 50, 100, 200, 500, and 1000 expert-validated labels from Seattle $\mathcal{D}_g$. The pre-trained FT-Transformer is fine-tuned on the equivalent number of expert-validated downstream samples.}
  \label{fig:performance}
\end{figure}

The results demonstrate that our pipeline enhances the efficiency of neural network training. The integration of relevant domain knowledge through PWS bypasses the resource-intensive task of manually labeling and validating the unannotated Project Sidewalk repository. As shown in \autoref{fig:performance}, our proposed pipeline with as few as 50 expert-validated labels (10 per label type), achieved a test accuracy and precision of 73.3\% and 83.6\%, respectively. It outperformed all baselines of traditional ML methods, even when those were trained on substantially more expert-validated labels. Our pipeline improved accuracy by up to 36.7\%, 28.0\%, 15.3\%, 16.3\%, 16.5\% for |$\mathcal{D}_g$| = 50, 100, 250, 500, 1000, respectively. Our pipeline also demonstrated an average 0.0859 boost in F1 score compared to the second-best method.

\subsubsection{Generalizability Across Cities}
As shown in \autoref{tab:chicago-oradell}, in Oradell, the pre-trained model, without fine-tuning, achieved an accuracy, precision, and recall of 80.8\%, 91.9\%, and 86.9\%, respectively. After the fine-tuning process, these figures rose to 91.4\%, 92.4\%, and 96.8\%. Similarly, for Chicago, the pre-trained model achieved accuracy, precision, and recall values of 67.9\%, 80.4\%, and 77.0\%, respectively. Post fine-tuning, these metrics improved to 71.9\%, 82.8\%, and 80.1\%. One plausible explanation for this lower performance could be the non-continuous geographic distribution of Project Sidewalk's data in Chicago, which includes a mix of dense urban downtown areas and pockets of suburbia. Variations in road width and intersection distances across these areas could complicate the model's ability to make accurate inferences.

\begin{table}[t]
\resizebox{\columnwidth}{!}{%
\begin{tabular}{@{}lll|ll@{}}
\toprule
            & \multicolumn{2}{c|}{Chicago} & \multicolumn{2}{c}{Oradell}   \\ \midrule
            & accuracy & F1 & accuracy & F1 \\
\rowcolor[HTML]{F3F3F3} 
Pre-trained on multi-city' $\mathcal{D}_n$    & 0.679 & 0.787 & 0.808 & 0.893  \\
Fine-tuned on target city's $\mathcal{D}_g$  & \textbf{0.719} & \textbf{0.814} & \textbf{0.914} & \textbf{0.945}  \\ \bottomrule

\end{tabular}%
}
\vspace{0.3cm}
\caption{Improvements of inference model in accuracy and F1 score for Chicago and Oradell after city-specific fine-tuning (fold K = 5).}
\label{tab:chicago-oradell}
\vspace{-0.75cm}
\end{table}

\subsubsection{Generalizability in New City}
To fully evaluate the model's generalizability, we deployed it to a new city: Newberg, OR—a small town in the Portland metropolitan area with a population of 25k, similar in urban composition to Oradell. When applying the pre-trained model to Newberg, which was neither previously pre-trained nor fine-tuned, the model showcased accuracy, precision, and recall of 78.3\%, 88.2\%, and 86.0\%, respectively. These scores represent on-par performance with the cities in the pre-training set, such as Oradell, NJ. This not only underscores the robust generalizable foundation of the multi-city pre-trained inference model, but also highlights that the pre-trained model can be deployed in a new city without any manual intervention and achieve respectable performance if the new city has a similar urban composition and crowdworker behavior to those in the pre-training set.

\subsubsection{Performance by Label Type}
We also analyzed performance as a function of label type for each city-specific fine-tuned model (\autoref{tab:city-comparison}). The model performs best for \textit{Curb Ramp} and \textit{Missing Sidewalk} across all cities, followed by \textit{Surface Problem}. However, \textit{Obstacle} is a low performer, especially in Seattle and Chicago. 
A close look at the tags associated with \textit{Obstacle} labels revealed that the observed discrepancies might be explained by the complexity of sidewalk obstacles in these two cities.
Specifically, in Oradell, obstacles were primarily associated with trees/vegetation (40\%), whereas in Seattle, \textit{Obstacle} were tagged with \textit{poles}, \textit{trash/recycling cans}, \textit{vegetation}, and \textit{parked cars} in similar frequencies of \textasciitilde20\%. Another low performer in Chicago is \textit{Missing Curb Ramp}, with a low F1 score of 0.494. This is associated with user behavior exclusive to Chicago, where curb ramps lacking tactile strips are often mislabeled as \textit{Missing Curb Ramp}. These findings highlight the high performance of our inference model for most scenarios, however further refinement is necessary to accommodate different urban environments and user behaviors.

\begin{table}[t]
\resizebox{\columnwidth}{!}{%
\begin{tabular}{@{}lrrrrr@{}}
\toprule
\rowcolor[HTML]{FFFFFF} 
City & 
\multicolumn{1}{c}{\cellcolor[HTML]{FFFFFF}\begin{tabular}[c]{@{}c@{}}Curb\\ Ramp\end{tabular}} & 
\multicolumn{1}{c}{\cellcolor[HTML]{FFFFFF}\begin{tabular}[c]{@{}c@{}} Missing \\ Curb Ramp\end{tabular}} & \multicolumn{1}{l}{\cellcolor[HTML]{FFFFFF}Obstacle} &
\multicolumn{1}{c}{\cellcolor[HTML]{FFFFFF}\begin{tabular}[c]{@{}c@{}}Surface\\ Problem\end{tabular}} & 
\multicolumn{1}{c}{\cellcolor[HTML]{FFFFFF}\begin{tabular}[c]{@{}c@{}}Missing\\ Sidewalk\end{tabular}} \\ \midrule
\rowcolor[HTML]{F3F3F3} 
Seattle       & 0.971                                                          & 0.966                                                                  & {\color[HTML]{FF6D01} \textbf{0.766}}                         & 0.861                                                                & 0.942                                                                 \\
\rowcolor[HTML]{FFFFFF} 
Chicago       & 0.968                                                          & {\color[HTML]{E06666} \textbf{0.494}}                                  & {\color[HTML]{FF6D01} \textbf{0.693}}                         & 0.718                                                                & 0.929                                                                 \\
\rowcolor[HTML]{F3F3F3} 
Oradell       & 0.972                                                          & 0.768                                                                  & 0.793                                                         & 0.944                                                                & 0.988                                                                 \\ \bottomrule
\end{tabular}
}
\vspace{0.3cm}
\caption{Performance by label type of our inference model in F1 score for Seattle, Chicago, and Oradell. \textit{Missing Curb Ramp} is a notable area of difficulty in Chicago. \textit{Obstacle} is a low performer in Seattle and Chicago.}
\label{tab:city-comparison}
\vspace{-0.25cm}
\end{table}

\begin{table*}[t]
\begin{tabular}{@{}llrlrlrlrlr@{}}
\toprule
\#Rank & \multicolumn{2}{l}{Curb Ramp} & \multicolumn{2}{l}{Missing Curb Ramp} & \multicolumn{2}{l}{Obstacle} & \multicolumn{2}{l}{Surface Problem} & \multicolumn{2}{l}{Missing Sidewalk} \\ \midrule
\rowcolor[HTML]{F3F3F3} 
1    & distance\_i      & 0.173     & distance\_i       & 0.177      & clustered        & 0.145     & zoom                & 0.143        & severity          & 0.129      \\
2    & way\_type        & 0.103     & way\_type         & 0.106      & zoom             & 0.135     & clustered           & 0.135        & tag               & 0.116      \\
\rowcolor[HTML]{F3F3F3} 
3    & severity         & 0.103     & tag               & 0.103      & description      & 0.134     & description         & 0.132        & distance\_r       & 0.101      \\ \bottomrule
\end{tabular}
\vspace{0.3cm}
\caption{Top 3 features and their importance coefficient per label type in Seattle. Note: distance\_i is the distance to intersection, distance\_r is the distance to road, and way\_type is the road hierarchy according to OpenStreetMap.}
\label{tab:feature-importance}
\end{table*}

\subsubsection{Feature Importance}
To explore feature importance, we used attention maps \cite{gorishniy_revisiting_2021}. We expect that if certain features are used as input variables in LFs for a specific sidewalk label type, these features will be highly significant in the model’s inference for that label type. For instance, the feature "distance to intersection" is an input feature for our model and also a variable in the LF ($\min \text{distance}(l, I)$ in Algorithm 1) for \textit{Curb Ramp} and \textit{Missing Curb Ramp}. 

The feature importance results of the model fine-tuned on Seattle $\mathcal{D}_c$ are shown in~\autoref{tab:feature-importance}. 
For label types \textit{Curb Ramp} and \textit{Missing Curb Ramp}, “distance to intersection" is the most important; while for \textit{Obstacle} and \textit{Surface Problem}, features like “zoom”, “cluster” and “description” are more crucial. 
This difference suggests that the features influencing \textit{Curb Ramps} are related to LFs based on urban planning knowledge, whereas those affecting \textit{Obstacles} and \textit{Surface Problems} are tied to user behavior. 
Specifically, mislabeled \textit{Curb Ramps} exhibit a spatial pattern, making them identifiable using domain knowledge (\textit{e.g.} \autoref{algorithm:algorithm-1}). 
In contrast, \textit{Obstacle} and \textit{Surface Problem} labeling mistakes are less about spatial distribution and more about user labeling diligence, characterized by zooming in and adding optional descriptions. The results of our feature importance ranking show how LFs based on urban planning and user behavior are complementary in achieving LabelAId’s performance. 
We believe such techniques generalize to other crowdsourcing platforms where user mistakes can be identified through a combination of domain guidelines as well as platform specific behaviors.

\begin{figure}[b]
  \centering
  \includegraphics[width=\linewidth]{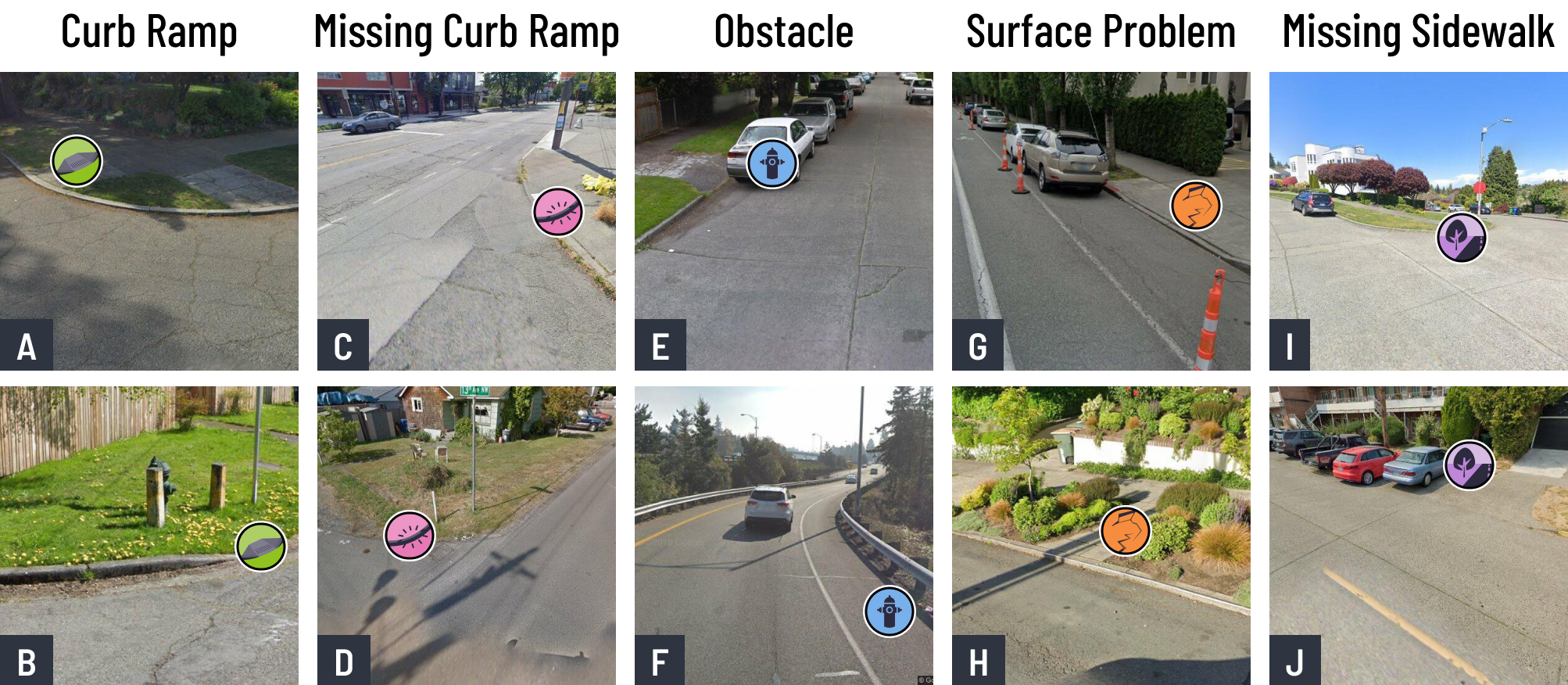}
  \caption{Selected typical inference false positives per label type (the actual label is wrong but was inferred as correct). a, c, failed to differentiate between a \textit{Curb Ramp} and a \textit{Missing Curb Ramp}. b, labeled a drainage swale near an intersection as \textit{Curb Ramp}. d, labeled \textit{Missing Curb Ramp} where there is no sidewalk. e-j, label has attributes for a correct label but there is ample space for a wheelchair user to pass.}
  \Description{Selected typical inference false positives per label type, where the model incorrectly infers the label as correct when, in fact, the label is wrong. a, c, failed to differentiate between a curb ramp and a missing curb ramp. b, d, labeled a drainage swale near an intersection with no sidewalk. For e to j, labels has attributes for a correct label but there is ample space for a wheelchair user to pass.}
  \label{fig:mistake-wrong-infer-as-correct}
\end{figure}

\subsubsection{Inference Limitations}
\begin{figure}[b]
  \centering
  \includegraphics[width=\linewidth]{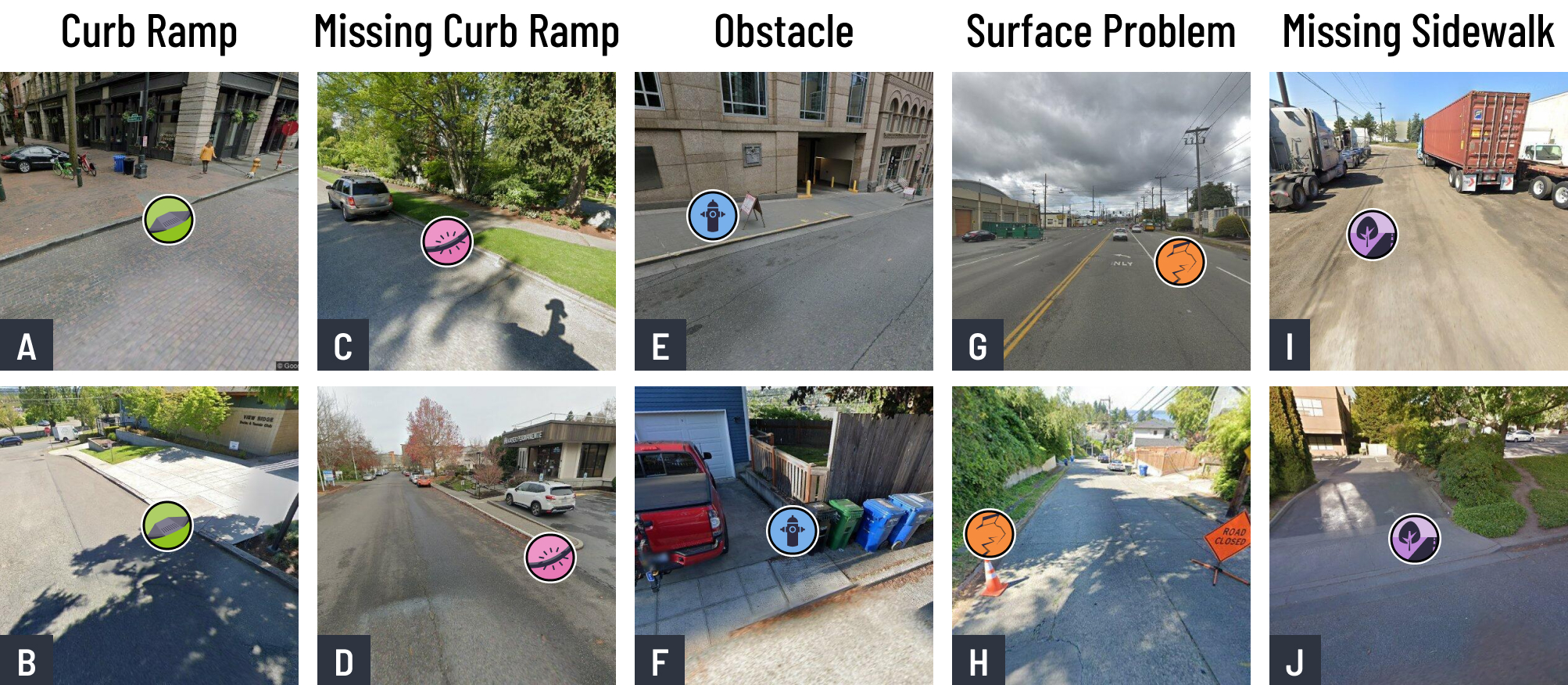}
  \caption{Selected typical inference false negatives per label type (the actual label is correct but was inferred as wrong). a-c, labeled mid-block crossings, geospatial information for such footpaths/crossings are incomplete in \textit{OpenStreetMap}. d, an exit for a public facility. e-h, missing optional inputs. i, j, rated \textit{Missing Sidewalk} with low severity.}
  \Description{Selected typical inference false negatives per label type, where the model incorrectly infers the label as wrong when, in fact, the label is correct. a to c are labeled mid-block crossings, geospatial information for such footpaths/crossings are incomplete in OpenStreetMap. d is an exit for a public facility. For e to h, missing optional inputs. i and j are labels rated Missing Sidewalk with low severity.}
  \label{fig:mistake-correct-infer-as-wrong}
\end{figure}

Finally, to understand the limitations of our model and identify opportunities for improvement, we conducted a qualitative assessment of our inference model by manually reviewing 100 randomly selected false positives and false negatives across each label type, and presented the results in \autoref{fig:mistake-wrong-infer-as-correct} \& ~\ref{fig:mistake-correct-infer-as-wrong}. 

In analyzing false positives (where the model incorrectly infers the label as correct when, in fact, the label is wrong), we observed two key sources of error for \textit{Curb Ramp} and \textit{Missing Curb Ramp}: (1) The model fails to differentiate between a \textit{Curb Ramp} and a \textit{Missing Curb Ramp }(\autoref{fig:mistake-wrong-infer-as-correct}a, c). (2) In edge cases, users labeled a drainage swale near an intersection as \textit{Curb Ramp} (\autoref{fig:mistake-wrong-infer-as-correct}b), and \textit{Missing Curb Ramp} where there was no sidewalk present (\autoref{fig:mistake-wrong-infer-as-correct}d). For \textit{Obstacle}, \textit{Surface Problem}, and \textit{Missing Sidewalk}, misclassifications typically occurred when a user's label included attributes for a correct label, but there was in fact ample space for wheelchair users to avoid the problem (\autoref{fig:mistake-wrong-infer-as-correct}e-j).
For false negatives, common sources of errors for \textit{Curb Ramp} and \textit{Missing Curb Ramp} included: (1) When users labeled mid-block crossings, geospatial information for such footpaths/crossings are incomplete in OpenStreetMap, causing inaccuracies when computing the distance to the nearest intersection (\autoref{fig:mistake-correct-infer-as-wrong}a-c). (2) The model struggled to correctly classify rare cases such as an exit for a public facility (\autoref{fig:mistake-correct-infer-as-wrong}d). For \textit{Obstacle}, and \textit{Surface Problem}, misclassifications happened when the problem could be easily identified without zooming in, thus contradicting the hypothesis that labels placed without zooming in are likely to be incorrect. Similar mistakes were found when the labels lacked inputs of tags, severity, and description—all of which are signals for a diligent crowdworker who typically produces more accurate labels (\autoref{fig:mistake-correct-infer-as-wrong}e-h). For \textit{Missing Sidewalk}, misclassifications often occurred when a user's label had a low severity rating, since the absence of sidewalks is supposed to be a high-severity issue (\autoref{fig:mistake-correct-infer-as-wrong}i, j).We refrain from further tuning of parameters in LFs post-analysis to prevent overfitting to specific scenarios in testing sets the model failed to learn, thereby preserving the model's generalizability.

\begin{figure*}[t]
  \centering
  \includegraphics[width=\linewidth]{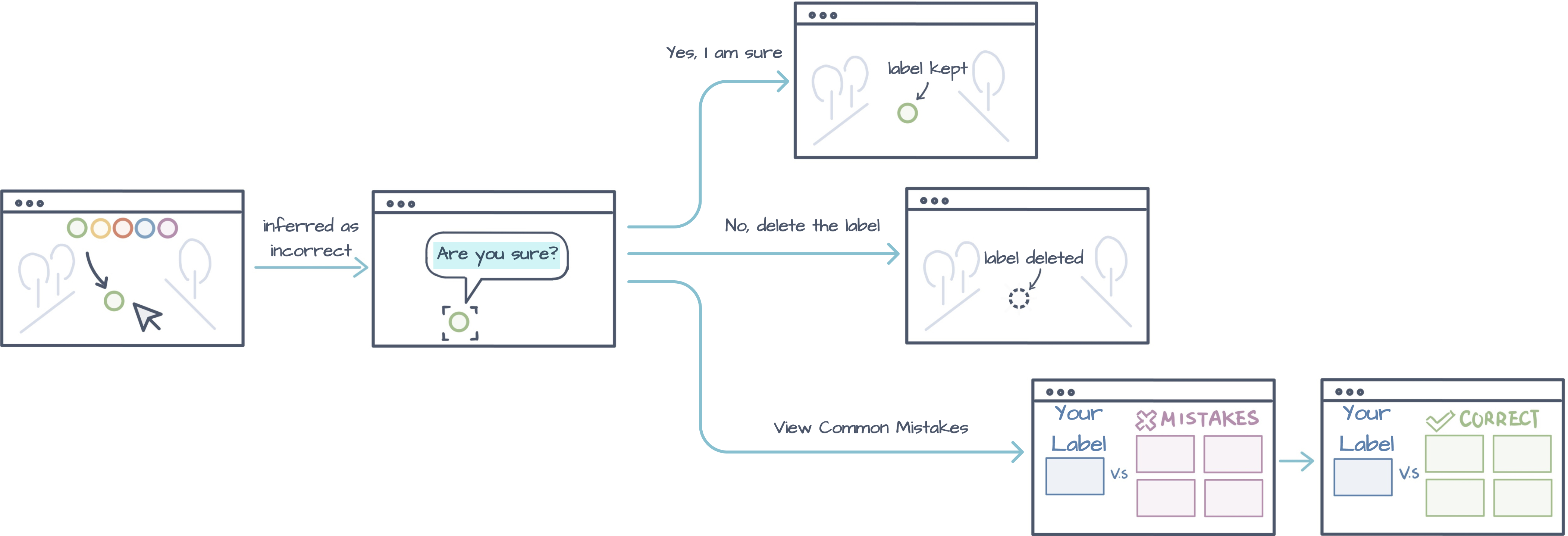}
  \caption{A user flow diagram of LabelAId implemented in Project Sidewalk. (1) A user places a label using the Project Sidewalk interface. (2) If LabelAId detects a mistake, the system displays a just-in-time intervention dialog. (3) The user can choose to keep the label, delete the label, or opt to view common mistakes associated with that label type. From the "View Common Mistakes" page, the user can navigate to the "View Correct Examples" page. See~\autoref{fig:system-interface} for actual screenshots.}
  \Description{This Figure illustrates the user flow diagram of LabelAId when implemented in Project Sidewalk. The process starts with a user placing a label using the Project Sidewalk interface. If LabelAId detects an error, a just-in-time intervention dialog appears. The user then has three options: keep the label, delete the label, or view common mistakes related to that label type. If the user chooses to view common mistakes, they can also navigate to a page showing correct examples.}
  \label{fig:user-flow}
\end{figure*}

\section{LabelAId: Implementation \& User Evaluation}\label{sec:user-eval}

Having demonstrated the technical efficacy of our LabelAId system in inferring label correctness, we implemented the LabelAId inference model in Project Sidewalk, and evaluated the user experience and performance of the end-to-end system with users in the loop. Our study aimed to answer the following questions:

\begin{itemize}
    \item RQ1: Can LabelAId's feedback improve the \textbf{\textit{performance}} of minimally-trained crowdworkers in labeling urban accessibility issues compared to a no feedback condition?
    \item RQ2: Can LabelAId's feedback enhance minimally-trained crowdworkers' \textbf{\textit{self-efficacy}} and \textbf{\textit{perceived learning}} when labeling urban accessibility issues compared to a no feedback condition?
    \item RQ3: How do participants perceive LabelAId's feedback in terms of \textbf{\textit{usefulness, content, and frequency}}?
\end{itemize}

To address these questions, we designed and conducted a between-subjects study of our LabelAId implementation, described below.

\subsection{Implementing LabelAId in Project Sidewalk}
To incorporate LabelAId into Project Sidewalk, we needed to integrate a real-time mistake inference model (as described in ~\autoref{sec:labelaid}) and to design and develop a just-in-time UI intervention to help warn users of potential labeling mistakes (using the said inference model). We first highlight design considerations situated in the literature, before describing implementation details.

\begin{figure*}[t]
  \centering
  \includegraphics[width=\linewidth]{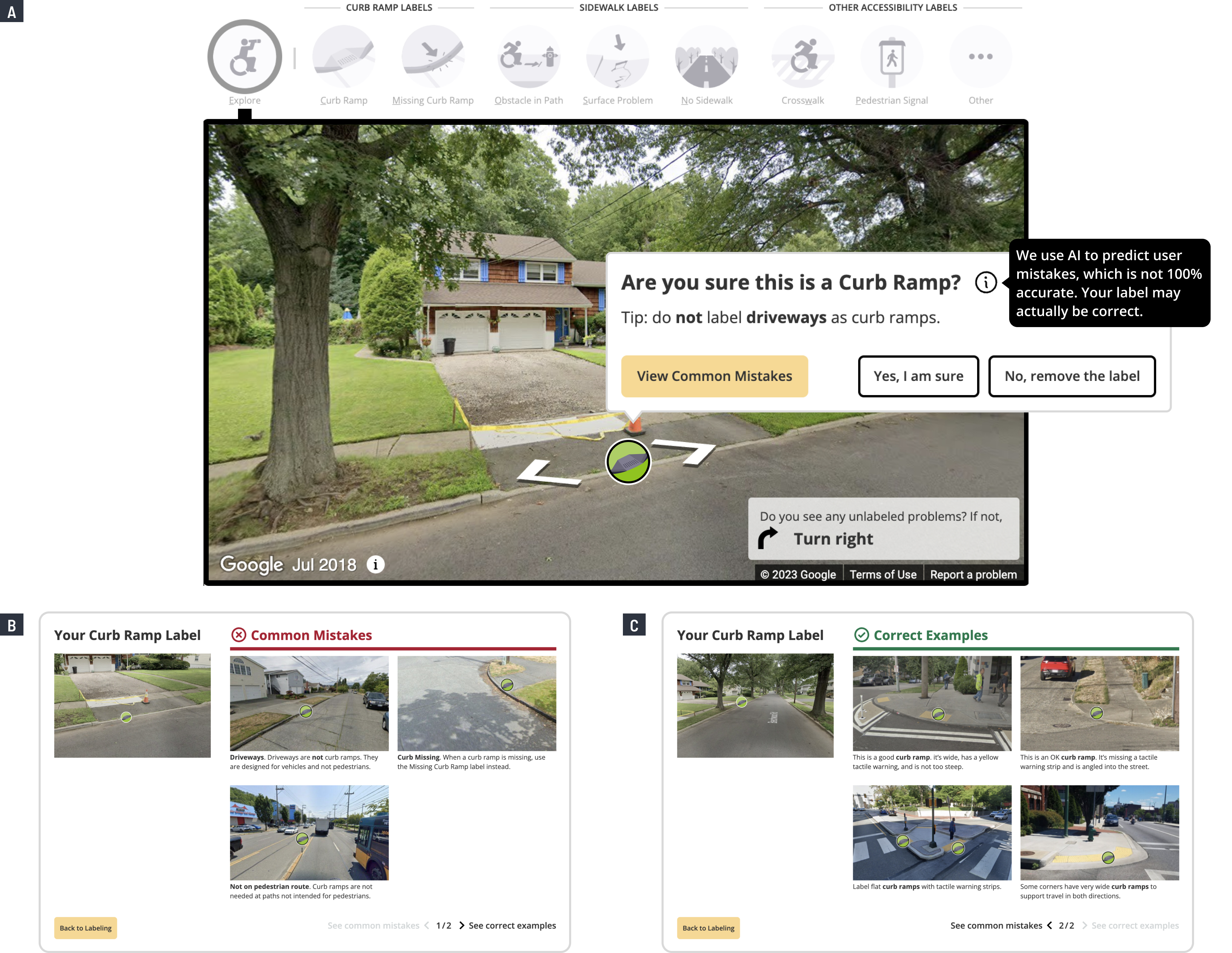}
  \caption{System screenshots of LabelAId implemented in Project Sidewalk. (A) When detecting a user label error: LabelAId pops up a just-in-time intervention dialog composed of three parts: a mistake title, a rotating set of labeling tips for that label type, and three buttons. "Yes, I am sure," "No, remove the label" or "View Common Mistakes". Hovering over the "i" icon will display a note explaining how the reminder is powered by AI and the system may make mistakes. (B) Common Mistakes Page. (C) Correct Examples Page. Both (B) and (C) present a screen capture of the user’s current label alongside three to four example labels, facilitating more straightforward comparison. }
  \Description{This figure contains system screenshots of LabelAId as implemented in Project Sidewalk. Part A shows a pop-up dialog that appears when a user label error is detected, featuring a mistake title, rotating labeling tips, and three action buttons: "Yes, I am sure," "No, remove the label," and "View Common Mistakes." Parts B and C are screenshots of the Common Mistakes Page and Correct Examples Page, respectively. Both pages display the user's current label alongside 3-4 example labels for easy comparison.}
  \label{fig:system-interface}
\end{figure*}

\textbf{Design considerations.} 
To design LabelAId's UI intervention, we first reviewed literature regarding the design space for crowd feedback~\cite{dow_shepherding_2012, zhu_reviewing_2014, wang_exploring_2018, doroudi_toward_2016, mamykina_learning_2016} and guidelines for HAI design~\cite{amershi_guidelines_2019}. Studies have emphasized the importance of timeliness in feedback delivery~\cite{dow_shepherding_2012}, which led us to opt for real-time feedback, as it delivers feedback during a \textit{teachable moment} when people are still thinking about the task. Additionally, the importance of contextual help for learning assistance has been well-documented in psychology literature~\cite{anderson_cognitive_1984} and demonstrated through HCI work (\textit{e.g.}, ~\cite{grossman_toolclips_2010, yeh_creating_2011}). To further refine the user interface, we consulted best practices for dialog design~\cite{nielsen_confirmation_2018}, emphasizing specific response options that clearly outline the consequences of each choice, as well as employing progressive disclosure techniques~\cite{nielsen_progressive_2006} to help users understand the implications of their actions before committing to them~\cite{amershi_guidelines_2019}. Based on these insights, we iteratively designed LabelAId, starting with hand sketches and \textit{Figma} mock-ups before implementing the tool in \textit{JavaScript} (front-end) and \textit{Scala} with \textit{QGIS} (back-end). 

\textbf{System implementation.} We integrated the city-specific, fine-tuned FT-Transformer into LabelAId using the \textit{Open Neural Network Exchange} (ONNX) runtime standard.  An important objective is to reduce latency and facilitate seamless HAI collaboration.The most time consuming step in the preparation stage is to assess whether the label belongs to a pre-existing cluster. 
To expedite calculation time, we simplified by calculating the spatial haversine distance of the input to a pre-computed cluster centroid, maintaining a threshold consistent with the clustering algorithm at 10 meters. 
We found in off-line experiments that this approach was 8-20 times faster (speed varies based on label type) and a mere 1.6\% of labels (27 out of 1659) had a different clustering result.

We implemented the inference model on the front-end rather than server-side for the following reasons: (1) Latency: considering the small model size (\textasciitilde100 KB), inference can be performed locally in the user's browser, thereby avoiding communication with a remote server and network latency. (2) Privacy: we reduced potential user privacy concerns, as no data is transmitted to a remote server for processing. Notably, during the user study, we found an average preparation time of 1.5 $ms$ and an average model inference time of 1.7 $ms$ across various hardware and platforms.

\textbf{User flow.} Drawing on previous research on crowdworker feedback~\cite{dow_shepherding_2012,hettiachchi_challenge_2021}, HAI~\cite{amershi_guidelines_2019}, and UI design~\cite{nielsen_progressive_2006}, we provide a two-stage intervention. After a user places a label, if LabelAId infers a mistake, we pop-up a just-in-time intervention dialog (\autoref{fig:system-interface}A) composed of three parts: a mistake title, a rotating set of labeling tips for that label type (\textit{e.g.}, "Do not label driveways as curb ramps."; see~\autoref{fig:system-interface}A), and three buttons: "Yes, I am sure," "No, remove the label" or "View Common Mistakes". Hovering over the "i" icon beside the mistake title will display an explanation that the reminder system is powered by AI and may make mistakes.
If the user selects "View Common Mistake", they enter the second stage of customized information about common mistakes and correct examples for that label type. To minimize users’ cognitive load~\cite{budiu_working_2018}, both the "View Common Mistakes" and "View Correct Examples" screens present a screen capture of the user's current label alongside three to four example labels, facilitating more straightforward comparison. These example images are curated based on an analysis of frequent mistakes and effective labeling practices on Project Sidewalk. Our user flow (\figcolor{\autoref{fig:user-flow}}) prompts users to reflect on their labeling decisions and then educate them through examples, both of which have been proven to enhance crowdwork quality~\cite{dow_shepherding_2012, zhu_reviewing_2014}.

\subsection{Study Design}
To examine our research questions, we conducted a between-subjects study with and without LabelAId. Inspired by previous Project Sidewalk mapathons, the study sessions were conducted in groups via Zoom based on condition. While this setup differs from traditional crowdsourcing studies conducted on platforms like MTurk or Prolific, mapathons and other synchronous social data collection events are key methods for participant involvement in crowdsourced mapping projects like Project Sidewalk and OpenStreetMap\footnote{\href{https://www.openstreetmap.org/}{https://www.openstreetmap.org/}
}. For example, in Project Sidewalk's 
 18-month deployment in Oradell, NJ, two single-day mapathons contributed over 2,056 labels, accounting for 22\% of all labels~\cite{li_i_2024}.
 
 Prior to the actual study sessions, we conducted pilot studies with one participant for each condition, during which two researchers observed the participants' labeling behaviors in-person and screen-recorded the process for post-analysis. Based on insights from these pilot studies, we refined the moderation workflow.

For the actual study, two study moderators led six online sessions, three for each condition. Each session had five to seven participants and lasted between 90 and 120 minutes. 
The sessions were composed of three parts, and the moderator adhered to a script to ensure consistency. 
First, we provided a brief orientation of urban accessibility and disability, guided the participants through platform account registration, and asked the participants to finish Project Sidewalk’s standard \textasciitilde5-minute interactive tutorial. 
Second, participants labeled eight curated routes on Project Sidewalk; the routes were carefully chosen by the research team to ensure they included frequent sidewalk accessibility features and problems. Both groups labeled identical routes. Participants were asked to mute themselves during the labeling tasks, and any questions were addressed privately via Zoom chat or in a breakout room.
Although the intervention group had access to correct and incorrect examples through the LabelAId UI flow, both groups were shown illustrated tutorial screens in the beginning of each route, which is the standard Project Sidewalk UI (\autoref{fig:labeling-help}). Furthermore, all participants could refer to these examples as well as the \textit{How to Label} section on the platform during labeling (\autoref{fig:labeling-help}), a practice we observed in both groups during the pilot studies.
Third, after completing their routes, participants filled out a post-study questionnaire followed by a semi-structured group debriefing session. The debriefing sessions were video and audio recorded.
Please see the supplementary materials for our orientation slide deck and pre- and post-study questionnaires.

\begin{table*}[t]
\resizebox{\textwidth}{!}{%
\begin{tabular}{@{}>{\RaggedRight\arraybackslash}p{0.24\textwidth}r|>{\RaggedRight\arraybackslash}p{0.24\textwidth}r|>{\RaggedRight\arraybackslash}p{0.24\textwidth}r@{}}
\toprule
\textbf{Labeling confidence} & \textbf{Count} & \textbf{Helpful elements during the labeling process }& \textbf{Count} & \textbf{Future improvement ideas} & \textbf{Count} \\ \midrule
\rowcolor[HTML]{F3F3F3} 
Confidence varies across different label types & 10 & Pop-ups & 11$^{*}$ & Implementing AI labeling followed by human verification & 10 \\
Confident grows with the labeling process & 6 & Tutorial & 6 & Providing rationales/confidence levels for the pop-ups & 3 \\
\rowcolor[HTML]{F3F3F3} 
High confidence in label type but uncertainty in severity ratings & 4 & Hover-over images & 5 & Introducing practice quiz to pre-filter participants & 2 \\ 
Unsure of potential missed labels
& 1
&  &  & Option to disable AI-generated pop-ups & 1 \\ 
\bottomrule
\end{tabular}
}
\vspace{0.3cm}
\caption{During the study’s semi-structured group debriefing session, we asked participants (N=34) open-response questions about their confidence levels during the labeling task, what was most helpful during the labeling process, and ideas for future improvements. Participants were not required to answer all questions. We manually coded the participants' responses to identify themes. The count column indicates the number of participants who mentioned each theme. $^{*}$Note that only the intervention group (N=17) was shown the pop-ups.}
\label{tab:thematic-analysis}
\end{table*}

\subsection{Participant Recruitment}
For our user study, we recruited participants via university mailing lists and snowball sampling. 
Our study size of 34 participants was determined through a power analysis using \textit{G*Power}~\cite{faul_g_2007}, aiming for an effect size of 1 and a statistical power of 0.8. 
Participants were randomly assigned to either the control or the intervention group depending on their availability. Based on self-reported demographics, we had 21 participants aged 18-24 (12 in the control group), 11 aged 25-34 (5 in the control group), and 2 aged 35-44 (none in the control group); 18 women (10 in the control group), 15 men (6 in the control group), and 1 non-binary individual (1 in the control group). As for computer experience, 2 participants reported having basic skills, 4 had intermediate skills, and 28 considered themselves experts; these numbers were evenly split between the two groups. Before the study session, all participants were required to sign a research consent form and complete a pre-study questionnaire. Each participant was compensated at a rate of \$30 per hour for their participation.

\subsection{Evaluation Measures}
Our study had a dual focus of understanding the objective performance of LabelAId users compared to the baseline as well as to examine their subjective experiences. For our objective measures, we collected and examined: 
\begin{itemize}
    \item \textbf{Labeling precision.} The number of correct labels compared to the total number of labels, measuring the correctness of user input.
    \item \textbf{Labeling time.} Time for participants to complete the labeling tasks, recorded per each route.
    \item \textbf{Learning gain in urban accessibility.} We designed quiz questions that were included in both pre- and post-study questionnaires (see supplementary materials). Participants were shown four images for each of the five label types and were asked to select the correct ones. A sum score was calculated for all participants: each correct answer earned 1 point, and each incorrect answer was penalized with -1 point.
\end{itemize}

We also captured subjective measures through 5-point Likert scale questions:
\begin{itemize}
     \item  \textbf{Confidence in response.} \textit{e.g.}, ``How confident are you in labeling curb ramps?''
     \item \textbf{Self-efficacy gain.} \textit{e.g.} ``I feel more confident about identifying problems on sidewalks faced by people with disabilities.''
     \item \textbf{Perceived learning gains in urban accessibility.} \textit{e.g.} ``Participating in the study gave me more ideas to make sidewalks accessible for people with disabilities.''
     \item \textbf{Perceived usefulness.} \textit{e.g. }``I liked the pop-up prompts.''
    \item  \textbf{Perceived AI intervention.} ``I felt that an AI agent was watching my performance/helping me while I was labeling.''
\end{itemize}
Full list of questions can be found in our supplementary materials. 

\subsection{Analysis Approach}
To analyze our results, two researchers independently validated all participant labels (N=3,574). In cases of disagreements (N=74, IRR=0.98), a third researcher was consulted to reach a consensus. Validations were then used to calculate the precision of user input. For subjective measures captured through Likert scale questions, we mapped responses such as "Strongly disagree" to "Strongly agree" or "Not confident at all" to "Very confident" onto a numerical scale ranging from 1 to 5. We then use descriptive statistics to explore the dataset and to assess the participant performance across different conditions. Due to the between-subjects study and the distribution of the data, we use \textit{Mann-Whitney U} tests to compare label precision, labeling time, and Likert scale responses between the two groups~\cite{robertson_modern_2016}.
Additionally, both the debriefing sessions and the post-study questionnaire included open-ended questions to capture nuanced feedback about perceived learning experience, self-efficacy, and overall user experience. Our analysis for these responses focused on summarizing high-level themes. One researcher developed a set of themes through qualitative open coding~\cite{charmaz_constructing_2006} based on the video transcript and the questionnaire responses, then coded the responses according to the themes. Participant quotes have been slightly modified for concision, grammar, and anonymity.

\subsection{Results}
During the study, participants contributed a total of 3,574 labels, with 2,091 from the control group and 1,483 from the intervention group. A detailed breakdown of the labels' types and their correctness can be found in~\autoref{tab:label-distribution}. Our open-encoding process highlighted several key themes, as outlined in~\autoref{tab:thematic-analysis}. When asked what helped the participants to label, a majority of intervention participants mentioned the pop-up screens. Regarding labeling confidence, they reported that their confidence varied across different label types and generally increased as they progressed through the tasks. In terms of future improvements, many suggested implementing AI-assisted labeling followed by human verification. Below, we delve into an in-depth analysis that integrates both qualitative and quantitative evaluations to address each research question.

\begin{table}[b]
\resizebox{\columnwidth}{!}{%
\begin{tabular}{@{}lrrrrrr@{}}
\toprule
Label Type &
  \multicolumn{2}{r}{Correct} &
  \multicolumn{2}{r}{Incorrect} &
  \multicolumn{2}{r}{Total}\\ \midrule
 &
  \multicolumn{1}{r}{C} &
  \multicolumn{1}{r}{I} &
  \multicolumn{1}{r}{C} &
  \multicolumn{1}{r}{I} &
  \multicolumn{1}{r}{C} &
  \multicolumn{1}{r}{I}\\
\rowcolor[HTML]{F3F3F3} 
Curb Ramp       & 436  & 454  & 487 & 23  & 923  & 477  \\
Missing Curb Ramp     & 265  & 245  & 61  & 29  & 326  & 274  \\
\rowcolor[HTML]{F3F3F3} 
Obstacle       & 309  & 298  & 124 & 77  & 433  & 375  \\
Surface Problem & 243  & 249  & 55  & 26  & 298  & 275  \\
\rowcolor[HTML]{F3F3F3} 
Missing Sidewalk     & 94   & 72   & 17  & 10  & 111  & 82   \\
Overall        & 1347 & 1318 & 744 & 165 & 2091 & 1483 \\ \bottomrule
\end{tabular}%
}
\vspace{0.3cm}
\caption{Distribution of participants’ labels across all label types. \textit{C} stands for Control group and \textit{I} stands for Intervention group.}
\label{tab:label-distribution}
\end{table}

\subsubsection{Task Performance (RQ1)}We first seek to examine whether there are significant differences between groups in task performance and how intervention level correlates with labeling precision within the intervention group.

\begin{figure*}[t]
  \centering
  \includegraphics[width=\linewidth]{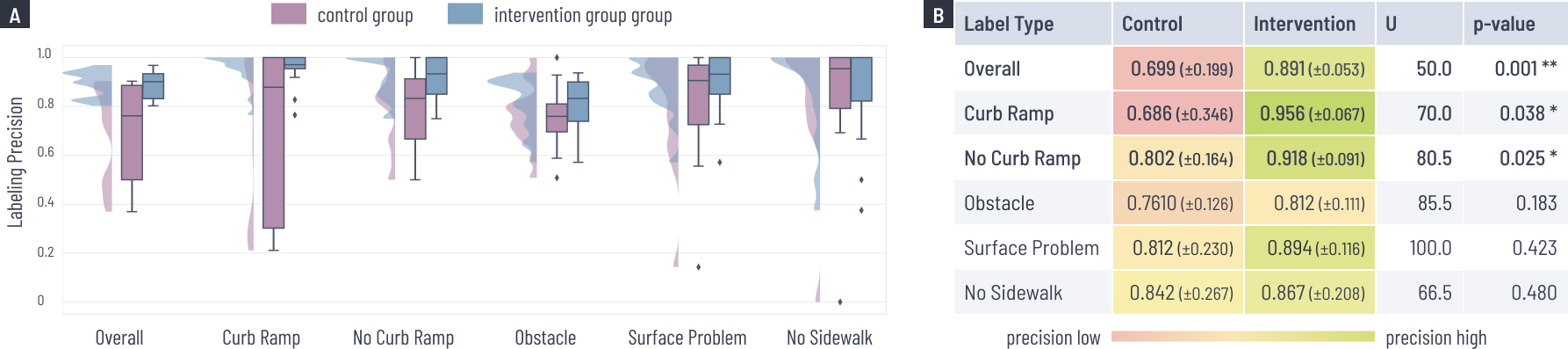}
  \caption{User labeling precision in the intervention group was higher across all categories, and the difference was statistically significant for the overall category, as well as for \textit{Curb Ramp }and \textit{Missing Curb Ramp} label types ($*p\leq$ 0.05, $**p \leq$ 0.01). (A) A raincloud plot (a half violin plot and a boxplot) shows user labeling precision between the control group and the intervention group, both overall and for the five specific label types. (B) A complementary table displays the precision mean, standard deviation, Mann-Whitney U value, and p-value for both the control and intervention groups.}
  \Description{This description presents data on user labeling precision. Part A features a raincloud plot, which combines a half violin plot and a box plot, to illustrate the difference in label precision between a control group and an intervention group. This is shown for overall precision and for five specific label types. Part B is a table that lists the mean, standard deviation, Mann-Whitney U value, and p-value for both groups. The table indicates that the intervention group had statistically significantly higher precision in the overall category, as well as for the "curb ramp" and "no curb ramp" label types.}
  \label{fig:results-precision-plot}
\end{figure*}

\textbf{Labeling precision and task completion time.} As summarized in~\autoref{fig:results-precision-plot}, the intervention group demonstrated higher precision overall and across all label types compared to the control group. The Mann-Whitney U results indicate a significant difference in precision between the two groups both overall ($p \leq$0.01) and for \textit{Curb Ramp} ($p \leq$0.05) and \textit{Missing Curb Ramp} ($p \leq$0.05) label types. For route completion time, we found no significant difference between the two groups ($p$=0.693). The control group had a mean completion time of 2303.3 seconds (SD=1240.3), while the intervention group spent 2801.4 seconds (SD=2035.3). Similarly, no significant differences were observed when examining the time taken for each of the eight routes (p-values ranged from 0.143 to 0.971). These findings indicate that the use of LabelAId resulted in improved labeling precision without compromising labeling speed.

\textbf{Labeling precision and level of intervention.} While the intervention group clearly performed better, two pertinent questions are: how \textbf{\textit{often}} did a LabelAId participant receive a just-in-time AI-assisted prompt and how \textbf{\textit{accurately}} did LabelAId perform, \textit{i.e.}, what was the true positive and false positive rate for intervening?

Towards examining the first question: within the intervention group, there were a total of 172 instances where LabelAId intervened with a just-in-time prompt (10.9\% of total labels; 10.1 per intervention group participant). When broken down by label type, LabelAId demonstrated high precision in predicting \textit{Curb Ramp} (0.882), \textit{Missing Curb Ramp} (0.750), and \textit{Missing Sidewalk} (1.000) mistakes. However, the model's precision was notably lower for \textit{Obstacle} (0.362) and \textit{Surface Problem} (0.377). Upon closer examination, we found that these less accurate inferences often corresponded with user behaviors that are likely to result in incorrect labels, such as not zooming in or failing to provide severity ratings or tags.

Within the 17 participants in the intervention group, our analysis revealed no significant correlation between the frequency of interventions by LabelAId and participants' labeling precision, either overall or for specific label types. Similarly, the number of times participants viewed common mistakes or correct examples UI screens did not correlate with their labeling accuracy (\autoref{tab:correlation}). We will return to this point in \autoref{sec:discussion}.

Despite the relatively low view frequency of the "Common Mistakes" UI screens (24 views in total, 1.4 views per person) and correct examples (6 views in total, 0.4 per person), qualitative feedback indicated their usefulness for those who chose to engage with them. During the debriefing sessions, several participants cited these screens when asked about what helped them during the labeling tasks. For instance, one participant noted a shift in their labeling approach after viewing the AI-triggered common mistake screen, stating, \textit{``Midway through, I saw the common mistakes, and it totally shifted my perspective. I had been labeling driveways from houses, but the screen clarified that those should not be labeled as curb cuts.''}

\subsubsection{Self-efficacy \& Learning Gains (RQ2)}
While the above findings demonstrate users' improvements in terms of task performance, we are also interested in self-efficacy and learning.

\textbf{Self-efficacy.} In the post-study questionnaire, we asked all participants about their confidence in identifying sidewalk features or problems. On average, participants rated their self-confidence higher in the intervention group (Avg=4.47; SD=0.88) than the control group (Avg=4.53; SD=0.52) with a statistically significant difference for \textit{Missing Curb Ramps} (Avg=4.6; SD=0.7 vs. Avg=3.8; SD=0.9, $p \leq$ 0.05), as shown in~\autoref{tab:five-label-types-confidence}. However, when participants were asked if they felt more confident about identifying problems on sidewalks faced by people with disabilities, the difference between groups was not statistically significant ($p$=0.721, see Q5 in ~\autoref{tab:questionnaire}).

\textbf{Perceived learning gains.} While task performance serves as one indicator of learning outcomes, we also used quizzes to assess objective learning gains and Likert scale questions to measure perceived learning gains. For objective learning gains, the mean improvement between the pre- and post-study quizzes was 1.35 (SD=1.73) for the control group and 1.31 (SD=1.54) for the intervention group, showing only a minor difference between the two. In terms of perceived learning gains, both groups demonstrated an enhanced understanding of curb ramps and accessibility challenges. Although the means were higher for the intervention group across all questions, no statistically significant difference was observed, except for the question, \textit{``Participating in the study gave me more ideas to make sidewalks accessible for people with disabilities.''}, where the mean score for the control group was 4.35 (SD=0.7), compared to 4.82 (SD=0.53) for the intervention group ($p \leq$0.05).

\subsubsection{Perceived Usefulness \& Presence of AI (RQ3)}
\label{sec:AI-usefulness}
Having explored the overall user performance, confidence and learning gain, we now turn to the perceived usefulness and presence of AI in LabelAId.

\textbf{Perceived usefulness.} Participants generally expressed a favorable view of LabelAId. When asked to what extent they agreed with the statements that the pop-up prompts were helpful and likable, the majority responded with "Somewhat Agree" or "Strongly Agree" (82.35\% and 64.7\%, respectively). In the post-study questionnaire and debriefing sessions, 11 out of 17 participants in the intervention group specifically cited the pop-up screens from LabelAId as a feature they appreciated or found helpful for labeling tasks. These timely reminders were particularly valued when participants were uncertain about their initial judgments. One participant mentioned: \textit{``There were times when I was not sure if I should label it, and the system popped-up for me and said ‘Are you sure about this?’ I found that really helpful.''} When asked about whether the prompts were distracting or appeared too frequently, the responses were more mixed—with a relatively even distribution across Likert responses.

\textbf{Perceived presence of AI.} We asked participants whether they felt an AI agent was observing their performance or assisting them during the labeling task and found a statistically significant differences between the two groups. This suggests that the presence of LabelAId had a noticeable impact on participants' perception of AI involvement. Interestingly, some participants in the control group explicitly expressed a desire for AI assistance. One control-group participant mentioned, \textit{``There was a section [in the post-study questionnaire] asking how I felt about AI helping me to label. Honestly, I didn't notice any AI while I was labeling. It would be super convenient if there was one that could suggest labels and ask me to correct them or provide a confidence level.''} This is exactly the intent of LabelAId.

\section{Discussion}
\label{sec:discussion}
Through our technical evaluation and user study, we showed how LabelAId improves both labeling data quality and crowdworkers' domain knowledge. 
We now situate our findings in related work, highlight key factors behind LabelAId's success, its limitations, and directions for future research. We also discuss how LabelAId can be generalized to other domains of crowdsourced science.

\subsection{Reflecting on LabelAId's Performance}
Below, we reflect on LabelAI'd performance and its relevance to future research, including comparing the differences between AI and human feedback, minimizing the overreliance on AI, and striking a balance between constructive feedback and perceived surveillance.

\textbf{Can AI-assistance replicate human-based feedback?} 
Prior work has shown that providing manual feedback to crowdworkers can improve task performance and enhance self-efficacy~\cite{dow_shepherding_2012, zhu_reviewing_2014, wang_exploring_2018, doroudi_toward_2016, mamykina_learning_2016}. Our study further reveals that AI-feedback can improve labeling performance, increase participants’ confidence, and enhance their domain knowledge—even with an imperfect ML inference model. 
While the nuances between human and AI-feedback in crowdsourcing have yet to be comprehensively studied, researchers in education have assessed the usage of automatic feedback as a learning tool~\cite{hegarty-kelly_analysis_2021,westera_automated_2018,hahn_systematic_2021,leite_effects_2020}.
Findings suggest that automatic feedback can reduce bias and increase consistency in grading~\cite{hegarty-kelly_analysis_2021}, liberate the instructor from grading to focus on other tasks~\cite{westera_automated_2018}, and allow more students to receive education simultaneously~\cite{singelmann_design_2019}. We believe that these benefits can well be extended to AI-generated feedback in crowdsourcing systems.

Yet, automated feedback in education contexts has limitations. It excels in grading tasks with clear-cut solutions (\textit{e.g.} programming questions), but may be challenging to implement in more subjective disciplines~\cite{hahn_systematic_2021}. Moreover, automatic graders fail to recognize when students are very close to meeting the criteria, whereas human graders would identify and assign partial grades accordingly~\cite{leite_effects_2020}. Future research in crowdsourcing should incorporate these insights from education science when designing AI-based feedback systems, and borrow approaches such as AI-feedback combined with human feedback on request~\cite{leite_effects_2020}.

\textbf{Cognitive forcing function reduces overreliance on AI.}
An overarching concern with AI-based assistance—including systems like LabelAId—is how the presence and behavior of  AI may actually \textit{reduce} active cognitive functioning in humans as they defer to AI's recommendations, which can then negatively impact overall task performance ~\cite{jacobs_how_2021,lai_human_2019}. For example, ~\cite{jacobs_how_2021,bucinca_proxy_2020} showed how users tend to overly depend on AI, following its suggestions even when their own judgment might be superior. Such a tendency is particularly problematic when the AI is inconsistent (\textit{e.g.,} across class categories), as in our case. 
 Recent work has explored \textit{cognitive forcing functions}~\cite{bucinca_trust_2021}—functions that elicit thinking at decision-making time. Because there is an \textit{anchoring bias}~\cite{green_principles_2019} that occurs when presenting users with AI's recommendations, one effective strategy is to ask the user to make a decision prior to seeing the AI's recommendation~\cite{bucinca_trust_2021}. Indeed, this is how LabelAId works: presenting suggestions only \textit{after} the user makes an initial decision and places a label—which may mitigate such bias.

Specifically, in our user study, LabelAId performed particularly poorly for two label types \textit{Obstacles} and \textit{Surface Problems} with false positive feedback rates of 36.2\% and 37.7\% respectively. 
However, users rejected these suggestions 83\% and 73\% of the time, indicating that they preferred their own judgments to the AI. Although this design choice was dictated by LabelAI’s model requirements, it encouraged analytical thinking that boosted participants' confidence in their own decisions. Our study contributes to the broader discourse of HAI, highlighting how system design can elicit analytical reasoning and reduce cognitive biases in decision-making.

\textbf{Striking a balance between constructive feedback and perceived surveillance.} 
We found a significant difference between the two groups regarding the perceived presence of AI (\Cref{sec:AI-usefulness}). 
Out of the 17 participants in the intervention group, eight felt observed and nine felt assisted by an AI agent, while in the control group, none felt observed and only three sensed AI assistance. 
We speculate that this difference in perceived surveillance also contributed to better intervention group performance, since they felt their work was being scrutinized.
This observation raises questions regarding AI agents as a form of surveillance in crowdsourcing environments.
When scholars apply a Foucauldian lens~\cite{foucault_discipline_1920} to monitoring technology, some see AI monitoring as social control from existing power hierarchies~\cite{calas_past_1999}, while others argue it can both restrict and empower individuals~\cite{lacombe_reforming_1996}.
This dichotomy implies that, if well-implemented, AI can encourage self-regulation among crowdworkers. 
A recent study confirms that digital feedback improves crowdwork outcomes when learning is the primary objective~\cite{wong_fostering_2021}, which is often the case in crowdsourcing in community science.
Therefore, we advocate for crowdsourcing platforms where the AI system strikes a balance between constructive feedback and perceived surveillance.

\subsection{LabelAId Limitations and Future Research} 
We now reflect on LabelAId's limitations and future work, focusing on designing interactions with imperfect ML models, promoting user agency in mixed-initiative interfaces, improving interaction efficiency in providing learning aids, and expanding participant diversity in future research.

\textbf{Designing interactions with imperfect ML models.}
With LabelAId, we were able to determine when a user likely made a mistake, but not the exact source of the error, which limited the types of prompting we could provide.
As one participant mentioned: \textit{"It’ll be great to provide some rationale or explanation on why there's a pop-up. Like maybe the location I placed my label is too far away from the obstacle.”} Current approaches of offering AI explainability falls into two categories: communicating information about the model inferences on a local level (\textit{e.g.} confidence score and local feature importance) and communicating information about the model itself on a global level (\textit{e.g.} model accuracy and global explanations)~\cite{lai_towards_2023}. However, LabelAId's current implementation does not incorporate explainability features.

On a global level, we recognize that our implementation could better communicate the model's varying accuracy levels across different label types. Despite a detailed technical analysis of LabelAId's performance in~\autoref{sec:labelaid}, we did not surface accuracy scores or global feature importance to participants. Future iterations should address this shortcoming. On a local level, we intentionally excluded confidence scores. This choice was informed by research indicating that confidence scores have limited impact on improving HAI collaboration~\cite{bansal_does_2021, bucinca_trust_2021}, coupled with our concern about over-cluttering the already busy UI. Future work may incorporate recent approaches to model the user's level of confidence and provide adaptive recommendations, \textit{i.e.,} only display AI's recommendations when the AI's confidence level is higher than the human's~\cite{ma_evaluating_2023}. 

In summary, while our current design decisions were informed by a balance of user cognitive load considerations and technical constraints, future work should explore other methods to provide users with tailored explanations and rationale, enhancing their understanding and interaction with the ML model. 

\textbf{Promoting user agency in mixed-initiative interfaces.}
Participants had mixed opinions about the frequency of AI interventions, with some finding them distracting. One participant noted, ``\textit{Sometimes the pop-ups were too frequent, so it might be helpful to give the user the option to disable them.}'' In addition, we also noticed the diminishing returns of increased intervention. During the study, there is no significant correlation between the frequency of intervention and task performance (\autoref{tab:correlation}). One potential explanation is that users understood their mistakes after the first few interventions, thereby making fewer mistakes in subsequent tasks. 
These findings, consistent with learning science research demonstrating that additional exposure or intervention does not necessarily improve performance (known as the \textit{saturation effect}~\cite{hauptmann_primed_2002}), are also supported by ongoing HAI research exploring ways to enhance human agency in mixed-initiative interfaces~\cite{amershi_guidelines_2019,lai_towards_2023,shneiderman_human-centered_2020}.
In future iterations, we would like to explore offering users overall control to enable or disable AI, to provide adaptive suggestion frequency based on labeling rate, and to allow users to request AI assistance only when needed~\cite{bucinca_trust_2021}.

\textbf{Designing efficient UI for learning aids.}
In addition to a lack of correlation between how often participants viewed example screens and their performance levels (\autoref{tab:correlation}), we observed that common mistakes and correct examples were only viewed a total of 30 times–six of the 17 intervention participants never viewed either of the screens.
This could be due to the \textit{interaction cost}~\cite{budiu_interaction_2013}:
the common mistakes screen requires two clicks and the correct examples screen three.
While click count alone is not a meaningful metric~\cite{laubheimer_3-click_2019}, it is important to minimize interaction costs~\cite{budiu_interaction_2013} by making key information easily accessible. 
Future work should explore developing effective methods for presenting examples to crowdworkers while they are balancing high cognitive load tasks.

\textbf{Expanding participant diversity in future research.}
While our study size of 34 aligns with typical HCI between-subjects studies (\textit{e.g}, ~\cite{jung_use_2021,palani_interweave_2022}), it is on the lower end for crowdsourcing research~\cite{kittur_future_2013}. However, our study design choice facilitated in-depth interviews and focused analysis, allowing us to gather qualitative insights not typical in crowdsourcing studies. 
Participants were recruited through snowball sampling from the research team's contacts and university mailing lists, which may not represent the comprehensive user base of Project Sidewalk including disability advocates. In future studies, we aim to enhance the applicability of our findings by expanding our participant base.

\subsection{Generalizability to Other Domains} 
Our study demonstrates the effectiveness of LabelAId in a crowdsourcing tool for urban accessibility, yet, its generalizability remains an open question.
We believe there are two primary generalizable components: 
\begin{itemize}
    \item \textbf{LabelAId’s PWS based ML pipeline.} PWS does not require annotated data, it works on a set of LFs generalized from domain knowledge and user behavior. This is particularly useful for crowdsourced community science because it allows organizers to transform their expertise and heuristic into LFs, which can then programmatically label large quantities of data. It is also more cost-effective compared to traditional ML models, as LabelAId improves inference accuracy by 36.7\% with only 50 downstream data points.
    \item \textbf{LabelAId’s mistake intervention design.} LabelAId's\textit{ in-situ }intervention design is rooted in literature on crowd feedback and contextual assistance, and aligns with recent HAI research on using cognitive theories to reduce over reliance on AI. Its simple two-step formula can be easily replicated in other platforms.
\end{itemize}

We believe our technique is most applicable to areas that require domain expertise and contextual understanding, such as medical image labeling~\cite{radsch_labelling_2023,zhang_samdsk_2023}, galaxy classification~\cite{simpson_zooniverse_2014}, and wildlife categorization~\cite{berger-wolf_wildbook_2017}. For example, the crowdsourcing application \textit{iNaturalist} uses identification technology and taxonomic experts to assist people in identifying natural species, and it achieves the best results when combined with traditional field guides~\cite{unger_inaturalist_2021}. We envision these guides and knowledge from experts being translated into LFs in our pipeline, and with similar mistake intervention design, LabelAId can help iNaturalist users contribute data more effectively while learning more about biodiversity.

\section{Conclusion}
In conclusion, LabelAId offers a practical approach to improving both crowdsourced data quality and domain knowledge of crowdworkers. By using machine learning to provide real-time feedback, LabelAId reduces the need for extensive manual review while also helping workers learn throughout the crowdsourcing process. Our user study demonstrates that LabelAId can improve user label quality without sacrificing speed, thereby offering a scalable solution to enhance worker knowledge and label quality in crowdsourcing tasks. While our empirical results focused on the performance of LabelAId within the context of urban accessibility, our framework can be extended to other crowdsourcing platforms, such as agricultural image recognition, medical imagery labeling, and wildlife biology image categorization.

\begin{acks}
We thank the reviewers for their insightful comments, our study participants, academic writing advisor Sandy Kaplan, and the Allen School Computer Science Laboratory Group. This work was supported by NSF SCC-IRG \#212508, PacTrans, an Amazon Research Award, and a Google Research Scholar. Zhihan Zhang is supported by the University of Washington CEI Fellowship.
\end{acks}
% \pagebreak

\bibliographystyle{ACM-Reference-Format}
\bibliography{references}
% \bibliography{streetscape-cv}

\appendix

\nobalance

\section{Additional Discussions on LabelAId Pipeline}

\subsection{Labeling Functions}
\label{app:lf-additional}
LFs serve as flexible interfaces within the framework of PWS. We assess three aspects of each LF (\autoref{tab:lf-table}): coverage (the proportion of examples each LF annotates), overlap (the proportion of examples each LF annotates that another LF also labels), and conflict (the proportion of examples each LF annotates and annotated differently by another LF). We note that it is necessary to apply as many LFs as possible for the best model performance due to the following reasons: (1) Improve coverage: each LF could capture different features of the data. More LFs can cover a higher proportion of raw data instances, leading to a larger AIA dataset generated from the PWS pipeline. (2) Reduce bias and overfitting: More LFs representing various heuristics or data insights, can mitigate systemic errors by averaging out individual LF bias. Incorporating multiple expert opinions and knowledge sources helps avoid overfitting to specific patterns or anomalies in the data, therefore making the model more generalizable. We assess the model performance with all LFs used during the PWS pipeline, and when removing one LF, the results (\autoref{tab:compare-all-lf}) show that even removing one LF during the PWS pipeline tends to hurt the end model's performance.

\begin{table}[h]
\begin{tabular}{@{}lrrrr@{}}
\toprule
                     & Polarity & Coverage &Overlaps & Conflicts \\ \midrule
\rowcolor[HTML]{F3F3F3} 
distance\_i & {[}0{]}          & 0.032             & 0.017             & 0.017             \\
clustered   & {[}1{]}          & 0.383             & 0.252             & 0.033             \\
\rowcolor[HTML]{F3F3F3} 
severity    & {[}0, 1{]}       & 0.066             & 0.062             & 0.051             \\
zoom        & {[}0, 1{]}       & 0.479             & 0.288             & 0.071             \\
\rowcolor[HTML]{F3F3F3} 
tag         & {[}1{]}          & 0.324             & 0.210             & 0.037             \\
description & {[}1{]}          & 0.010             & 0.010             & 0.004             \\
\rowcolor[HTML]{F3F3F3} 
distance\_r & {[}0{]}          & 0.027             & 0.027             & 0.019             \\
way\_type   & {[}0, 1{]}       & 0.030             & 0.023             & 0.016             \\ \bottomrule
\end{tabular}
\caption{\rev{Labeling function analysis using label matrix. [0] = Wrong, [1] = Correct.Note: distance\_i is the distance to intersection, distance\_r is the distance to road, and way\_type is the road hierarchy according to OpenStreetMap.}}
\label{tab:lf-table}
\end{table}

\subsection{Programmatic Weak Supervision vs. Hard Rule-based Approach}
\label{sec:pws-additional}
A key aspect of PWS is its ability to handle noise and conflicts in LFs~\cite{ratner_snorkel_2017, ratner_training_2019, ratner_data_2016}. Hard rule-based approaches would struggle in scenarios where LFs conflict or where the data presents ambiguities. For instance, if a user places a \textit{Missing Curb Ramp} label within the distance threshold to the intersection but fails to provide a tag, then LFs of \textit{distance\_i} and \textit{tag} provide contradictory annotations. PWS integrates these imperfect LFs into a probabilistic graphical model, so it can evaluate these conflicts based on the learned weight of each LF, whereas a hard rule-based approach would lack the mechanism to resolve such conflicts.

Our analysis indicates that LabelAId outperforms a hard rule-based method across all five label types (\autoref{tab:compare-hard-rule}). 
To mitigate the complexity of resolving conflicts, we selected the most important rule from our feature importance analysis for each label type (\autoref{tab:feature-importance}).  
However, it is worth noting that hard rule-based approaches may still be valuable in low-resource scenarios. 
In situations where the raw dataset is small or when there is limited computational capacity to run an AI inference model, crafting a few expert-defined rules might be more feasible and efficient than establishing a complex PWS setup. 

\newpage
\begin{table}[h]
\centering
\resizebox{\columnwidth}{!}{
\begin{tabular}{@{}
    >{\raggedright\arraybackslash}p{\dimexpr 0.23\columnwidth-2\tabcolsep}
    >{\raggedright\arraybackslash}p{\dimexpr 0.13\columnwidth-2\tabcolsep}
    >{\raggedright\arraybackslash}p{\dimexpr 0.15\columnwidth-2\tabcolsep}
    >{\raggedright\arraybackslash}p{\dimexpr 0.15\columnwidth-2\tabcolsep}
    >{\raggedright\arraybackslash}p{\dimexpr 0.17\columnwidth-2\tabcolsep}
    >{\raggedright\arraybackslash}p{\dimexpr 0.18\columnwidth-2\tabcolsep}@{}}
\toprule
    & Curb Ramp & Missing Curb & Obstacle & Surface Problem & Missing Sidewalk \\ \midrule
\rowcolor[HTML]{F3F3F3} 
With all LFs & \textbf{0.971}    & \textbf{0.966}      & \textbf{0.766}    & \textbf{0.861}     & \textbf{0.942} \\
Without distance\_r & 0.909    & 0.918      & 0.695    & 0.803     & 0.836 \\ 
\rowcolor[HTML]{F3F3F3} 
Without tag & 0.910    & 0.885      & 0.556    & 0.677     & 0.722 \\ \bottomrule
\end{tabular}
}
\caption{Performance decreases by label type of our LabelAId pipeline in F1 score for Seattle after one LF being removed. Note: distance\_r is the distance to the road.}
\label{tab:compare-all-lf}

\centering
\resizebox{\columnwidth}{!}{
\begin{tabular}{@{}
    >{\raggedright\arraybackslash}p{\dimexpr 0.2\columnwidth-2\tabcolsep}
    >{\raggedright\arraybackslash}p{\dimexpr 0.15\columnwidth-2\tabcolsep}
    >{\raggedright\arraybackslash}p{\dimexpr 0.15\columnwidth-2\tabcolsep}
    >{\raggedright\arraybackslash}p{\dimexpr 0.15\columnwidth-2\tabcolsep}
    >{\raggedright\arraybackslash}p{\dimexpr 0.18\columnwidth-2\tabcolsep}
    >{\raggedright\arraybackslash}p{\dimexpr 0.18\columnwidth-2\tabcolsep}@{}}
\toprule
    & Curb Ramp & Missing Curb & Obstacle & Surface Problem & Missing Sidewalk \\ \midrule
\rowcolor[HTML]{F3F3F3} 
LabelAId & \textbf{0.971}    & \textbf{0.966}      & \textbf{0.766}    & \textbf{0.861}     & \textbf{0.942} \\
Hard Rule-based & 0.943    & 0.752      & 0.660    & 0.576     & 0.849 \\ \bottomrule
\end{tabular}
}
\caption{\rev{Performance by label type of our LabelAId pipeline compared to the hard rule-based approach in F1 score for Seattle.}}
\label{tab:compare-hard-rule}
\end{table}

\newpage
\section{User Evaluation Tables}
% Table 10
\begin{table}[h]
\begin{tabular}{@{}lrl@{}}
\toprule
\rowcolor[HTML]{FFFFFF} 
 &
  \multicolumn{1}{l}{\cellcolor[HTML]{FFFFFF}rho} &
  \multicolumn{1}{l}{\cellcolor[HTML]{FFFFFF}p-value} \\ \midrule
\rowcolor[HTML]{F3F3F3} 
\begin{tabular}[c]{@{}l@{}}Number of times being intervened\end{tabular} &
  -0.141 &
  0.589 \\
\rowcolor[HTML]{FFFFFF} 
\begin{tabular}[c]{@{}l@{}}Total time spent interacting with UI\end{tabular} &
  -0.230 &
  0.374 \\
\rowcolor[HTML]{F3F3F3} 
\begin{tabular}[c]{@{}l@{}}Times viewed common mistakes\end{tabular} &
  -0.066 &
  0.801 \\
\rowcolor[HTML]{FFFFFF} 
\begin{tabular}[c]{@{}l@{}}Times viewed correct examples\end{tabular} &
  -0.004 &
  0.989 \\
\rowcolor[HTML]{F3F3F3} 
\begin{tabular}[c]{@{}l@{}}Total times viewed example screens\end{tabular} &
  -0.074 &
  0.779 \\ 
  \bottomrule
\end{tabular}

\caption{Spearman's rho correlation results for the level of intervention and precision.}
\label{tab:correlation}

\begin{tabular}{@{}lrrrl@{}}
\toprule
Quiz &
  \multicolumn{1}{l}{Control} &
  \multicolumn{1}{l}{Intervention} &
  \multicolumn{1}{l}{U} &
  \multicolumn{1}{l}{p-value} \\ \midrule
\rowcolor[HTML]{F3F3F3} 
Pre-study  & 5.53 (2.07) & 5.06 (1.57) & 163.50 & 0.51 \\
% \rowcolor[HTML]{F3F3F3} 
Post-study & 6.88 (1.45) & 6.38 (2.09) & 174.00 & 0.30 \\
\rowcolor[HTML]{F3F3F3} 
Delta      & 1.35 (1.73) & 1.31 (1.54) & 152.50 & 0.78 \\ 
\bottomrule
\end{tabular}%
 \caption{Quiz scores. In both pre- and post-study questionnaires, participants were shown four images for each of the five label types and were asked to select the correct ones. A sum score was calculated for all participants: each correct answer earned 1 point, and each incorrect answer was penalized with -1 point. There was no statistical difference between the two groups.}
\label{tab:quiz-score}

\begin{tabular}{@{}lrrrl@{}}
\toprule
\rowcolor[HTML]{FFFFFF} 
Question                    & \multicolumn{1}{l}{\cellcolor[HTML]{FFFFFF}Control} & \multicolumn{1}{l}{\cellcolor[HTML]{FFFFFF}Intervention} & \multicolumn{1}{l}{\cellcolor[HTML]{FFFFFF}U} & \multicolumn{1}{l}{\cellcolor[HTML]{FFFFFF}p-value} \\ \midrule
\rowcolor[HTML]{F3F3F3} 
Curb Ramp                  & \cellcolor[HTML]{E6E295}4.65                        & \cellcolor[HTML]{DDE085}4.71                             & 142.0                                         & 0.914                                               \\
\rowcolor[HTML]{FFFFFF} 
\textbf{Missing Curb} & \cellcolor[HTML]{EFBAB5}\textbf{3.88}               & \cellcolor[HTML]{F0E5A6}\textbf{4.59}                    & \textbf{84.5}                                 & \textbf{0.023*}                                     \\
\rowcolor[HTML]{F3F3F3} 
Obstacles                   & \cellcolor[HTML]{F6DAB5}4.35                        & \cellcolor[HTML]{D3DD74}4.76                             & 98.0                                          & 0.061                                               \\
\rowcolor[HTML]{FFFFFF} 
Surface Problems            & \cellcolor[HTML]{F3CEB5}4.18                        & \cellcolor[HTML]{F8E2B5}4.47                             & 116.0                                         & 0.276                                               \\
\rowcolor[HTML]{F3F3F3} 
Missing Sidewalk            & \cellcolor[HTML]{F7DEB5}4.41                        & \cellcolor[HTML]{E6E295}4.65                             & 123.5                                         & 0.392                                               \\ \bottomrule
\end{tabular}
\caption{Responses to the question: "How confident are you that you can correctly recognize the following?"  We mapped responses from "Not confident at all" to "Very confident" to 1-5. (*p $\leq$ 0.05, **p $\leq$ 0.01, ***p $\leq$ 0 .001). }
\label{tab:five-label-types-confidence}
\end{table}

% Table 13
\begin{table*}[h] 
\begin{tabular}{@{}rp{9cm}rrrl@{}}
\toprule
\rowcolor[HTML]{FFFFFF} 
\multicolumn{1}{l}{\cellcolor[HTML]{FFFFFF}\#} & Question                                                                                                     & \multicolumn{1}{l}{\cellcolor[HTML]{FFFFFF}Control} & \multicolumn{1}{l}{\cellcolor[HTML]{FFFFFF}Intervention} & \multicolumn{1}{l}{\cellcolor[HTML]{FFFFFF}U} & p-value \\ \midrule
\rowcolor[HTML]{F3F3F3} 
1                                                       & I feel that I have a better understanding of what a sidewalk curb ramp (or curb cut) is.                              & \cellcolor[HTML]{E9E39A}4.59                                 & \cellcolor[HTML]{DEE087}4.71                                      & 127.5                                                  & 0.480            \\
\rowcolor[HTML]{FFFFFF} 
2                                                       & I feel that I understand better the accessibility challenges people with disabilities have to participate in society. & \cellcolor[HTML]{F4E6AD}4.47                                 & \cellcolor[HTML]{E9E39A}4.59                                      & 146.0                                                  & 0.952            \\
\rowcolor[HTML]{F3F3F3} 
3                                                       & I feel that I have a better understanding of the sidewalk barriers that impact people who use wheelchairs or walkers. & \cellcolor[HTML]{DEE087}4.71                                 & \cellcolor[HTML]{DEE087}4.71                                      & 150.5                                                  & 0.788            \\
\rowcolor[HTML]{FFFFFF} 
4                                                       & I feel that I have a better understanding of the sidewalk barriers that impact people who are blind or low-vision.    & \cellcolor[HTML]{F7DFB5}4.00                                 & \cellcolor[HTML]{F8E4B5}4.29                                      & 120.5                                                  & 0.373            \\
\rowcolor[HTML]{F3F3F3} 
5                                                       & I feel more confident about identifying  problems on sidewalks faced by people with disabilities                      & \cellcolor[HTML]{F4E6AD}4.47                                 & \cellcolor[HTML]{EFE5A4}4.53                                      & 153.5                                                  & 0.721            \\
\rowcolor[HTML]{FFFFFF} 
\textbf{6}                                              & \textbf{Participating in the study gave me more ideas to make sidewalks accessible for people with disabilities.}     & \cellcolor[HTML]{F8E5B5}\textbf{4.35}                        & \cellcolor[HTML]{D3DD74}\textbf{4.82}                             & \textbf{87.5}                                          & \textbf{0.017*}  \\
\rowcolor[HTML]{F3F3F3} 
7                                                       & I enjoyed using Project Sidewalk.                                                                                     & \cellcolor[HTML]{F8E3B5}4.24                                 & \cellcolor[HTML]{F4E6AD}4.47                                      & 119.5                                                  & 0.336            \\
\rowcolor[HTML]{FFFFFF} 
8                                                       & It was easy for me to use Project Sidewalk.                                                                           & \cellcolor[HTML]{F4E6AD}4.47                                 & \cellcolor[HTML]{F8E4B5}4.29                                      & 153.5                                                  & 0.728            \\
\rowcolor[HTML]{F3F3F3} 
\textbf{9}                                              & \textbf{I felt that an AI agent was watching my performance while I was labeling.}                                    & \cellcolor[HTML]{EFBAB5}\textbf{1.82}                        & \cellcolor[HTML]{F3CEB5}\textbf{3.00}                             & \textbf{68.0}                                          & \textbf{0.006**} \\
\rowcolor[HTML]{FFFFFF} 
\textbf{10}                                             & \textbf{I felt that an AI agent was helping me throughout the task.}                                                  & \cellcolor[HTML]{F0C2B5}\textbf{2.29}                        & \cellcolor[HTML]{F4D2B5}\textbf{3.24}                             & \textbf{83.5}                                          & \textbf{0.028*}  \\
\rowcolor[HTML]{F3F3F3} 
11                                                      & Overall, I desired more active help to complete the labeling tasks.                                                   & \cellcolor[HTML]{F6DDB5}3.88                                 & \cellcolor[HTML]{F4D2B5}3.24                                      & 189.0                                                  & 0.106            \\ \bottomrule
\end{tabular}
\caption{Responses to the question "To what extent do you agree with the following statements?". We mapped responses such as "Strongly disagree" to "Strongly agree" to 1-5. (*p $\leq$ 0.05, **p $\leq$ 0.01, ***p $\leq$ 0 .001).}
\label{tab:questionnaire}
\end{table*}

\begin{figure*}[b] 
  \section{Labeling Assistance Interface}
  \vspace{.6cm}
  \centering
  \includegraphics[width=\linewidth]{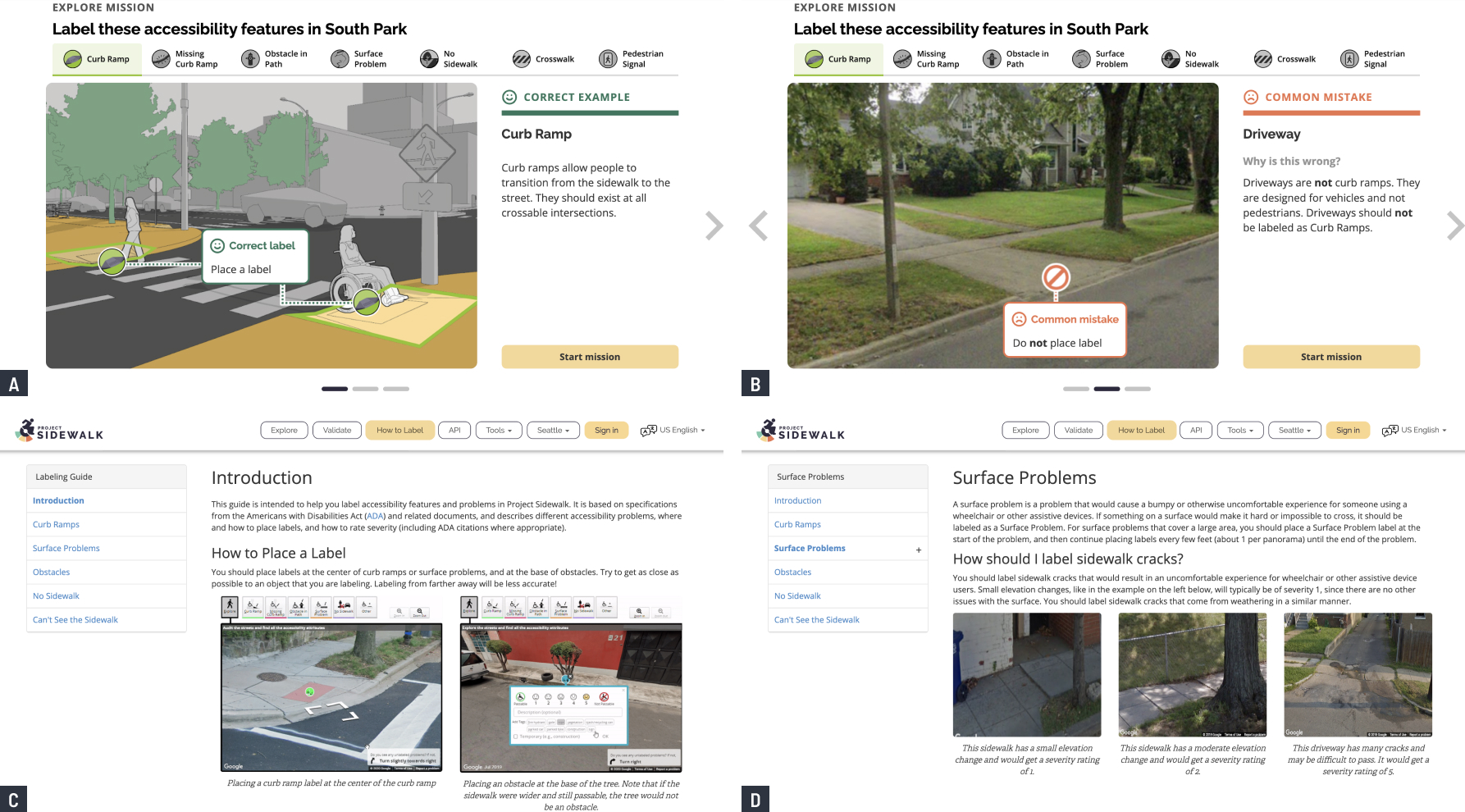}
  \caption{Project Sidewalk built-in labeling assistance. (A) \& (B) Illustrated example screens shown in the beginning of each route. The label type is rotated every time. (C) \& (D) The \textit{How to Label }Section. Participants may access this section at any time during the labeling process.}
  \Description{
  Four screenshots demonstrating the features of Project Sidewalk's built-in labeling assistance. Screenshots A and B display example screens presented at the start of each route, showing different types of labels that rotate each time. Screenshots C and D show the 'How to Label' section, which participants can access at any point during the labeling process to receive guidance and instructions.}
  \label{fig:labeling-help}
\end{figure*}

\end{document}